\documentclass[10pt,journal,compsoc]{IEEEtran}
\usepackage{amsmath,amsfonts}
\usepackage{algorithmic}
\usepackage{algorithm}
\usepackage{array}
\usepackage[table,xcdraw,dvipsnames,svgnames]{xcolor}
\usepackage{booktabs}  
\usepackage[caption=false,font=normalsize,labelfont=sf,textfont=sf]{subfig}
\usepackage{textcomp}
\usepackage{stfloats}
\usepackage{url}
\usepackage{verbatim}
\usepackage{graphicx}
\usepackage{orcidlink}
\usepackage{cite}
\usepackage{wrapfig}
\usepackage{mathptmx}                  % use matching math font
\usepackage{ragged2e}
\usepackage{tabu}
\usepackage{multirow}
\usepackage{soul}
\usepackage{tikz}
\usetikzlibrary{tikzmark, arrows.meta,calc}
\definecolor{revision}{RGB}{0,0,0}
\definecolor{revision2}{RGB}{0,0,0} % Milad's revisions
\newcommand{\revDoug}[1]{\textcolor{black}{#1}}
\newcommand{\revAli}[1]{\textcolor{black}{#1}} % Ali's revisions
\newcommand{\revMilad}[1]{\textcolor{black}{#1}}

\newcommand{\revDougFinal}[1]{\textcolor{black}{#1}}
\newcommand{\revAliFinal}[1]{\textcolor{black}{#1}}
\newcommand{\revMiladFinal}[1]{\textcolor{black}{#1}}

\graphicspath{{figs/}{figures/}{pictures/}{images/}{./}} % where to search for the images

\begin{document}

\title{The Impact of Elicitation and Contrasting Narratives on Engagement, Recall and Attitude Change with News Articles Containing Data Visualization}

\author{Milad Rogha\,\orcidlink{0000-0002-1464-2157},
  Subham Sah,
  Alireza Karduni\, \orcidlink{0000-0001-9719-7513},
  Douglas Markant, and
  Wenwen Dou
        % <-this % stops a space
\thanks{Milad Rogha, Subham Sah, and Wenwen Dou are with the Department of Computer Science, University of North Carolina at Charlotte.\\Doug Markant is a faculty at the Department of Psychological Science, University of North Carolina at Charlotte.\\ Alireza Karduni is is an Assistant Professor at Simon Fraser University’s School of Interactive Arts and Technology.}% <-this % stops a space
}

% The paper headers
% \markboth{Journal of \LaTeX\ Class Files,~Vol.~14, No.~8, August~2021}%
% {Shell \MakeLowercase{\textit{et al.}}: A Sample Article Using IEEEtran.cls for IEEE Journals}

% \IEEEpubid{0000--0000/00\$00.00~\copyright~2021 IEEE}
% Remember, if you use this you must call \IEEEpubidadjcol in the second
% column for its text to clear the IEEEpubid mark.

\maketitle

\begin{abstract}
News articles containing data visualizations play an important role in informing the public on issues ranging from public health to politics. 
Recent research on the persuasive appeal of data visualizations suggests that prior attitudes can be notoriously difficult to change. 
Inspired by an NYT article, we designed two experiments to evaluate the impact of elicitation and contrasting narratives on attitude change, recall, and engagement.
We hypothesized that eliciting prior beliefs leads to more elaborative thinking that ultimately results in higher attitude change, better recall, and engagement. Our findings revealed that visual elicitation leads to higher engagement in terms of feelings of surprise. While there is an overall attitude change across all experiment conditions, we did not observe a significant effect of belief elicitation on attitude change. With regard to recall error, while participants in the draw trend elicitation exhibited significantly lower recall error than participants in the categorize trend condition, we found no significant difference in recall error when comparing elicitation conditions to no elicitation. In a follow-up study, we added contrasting narratives with the purpose of making the main visualization (communicating data on the focal issue) appear strikingly different. Compared to the results of study 1, we found that contrasting narratives improved engagement in terms of surprise and interest but interestingly resulted in higher recall error and no significant change in attitude. We discuss the effects of elicitation and contrasting narratives in the context of topic involvement and the strengths of temporal trends encoded in the data visualization.
\end{abstract}


\section{Introduction}

\begin{wrapfigure}{l}{0.2\textwidth} %this figure will be at the right
    \centering
    \includegraphics[width=0.2\textwidth]{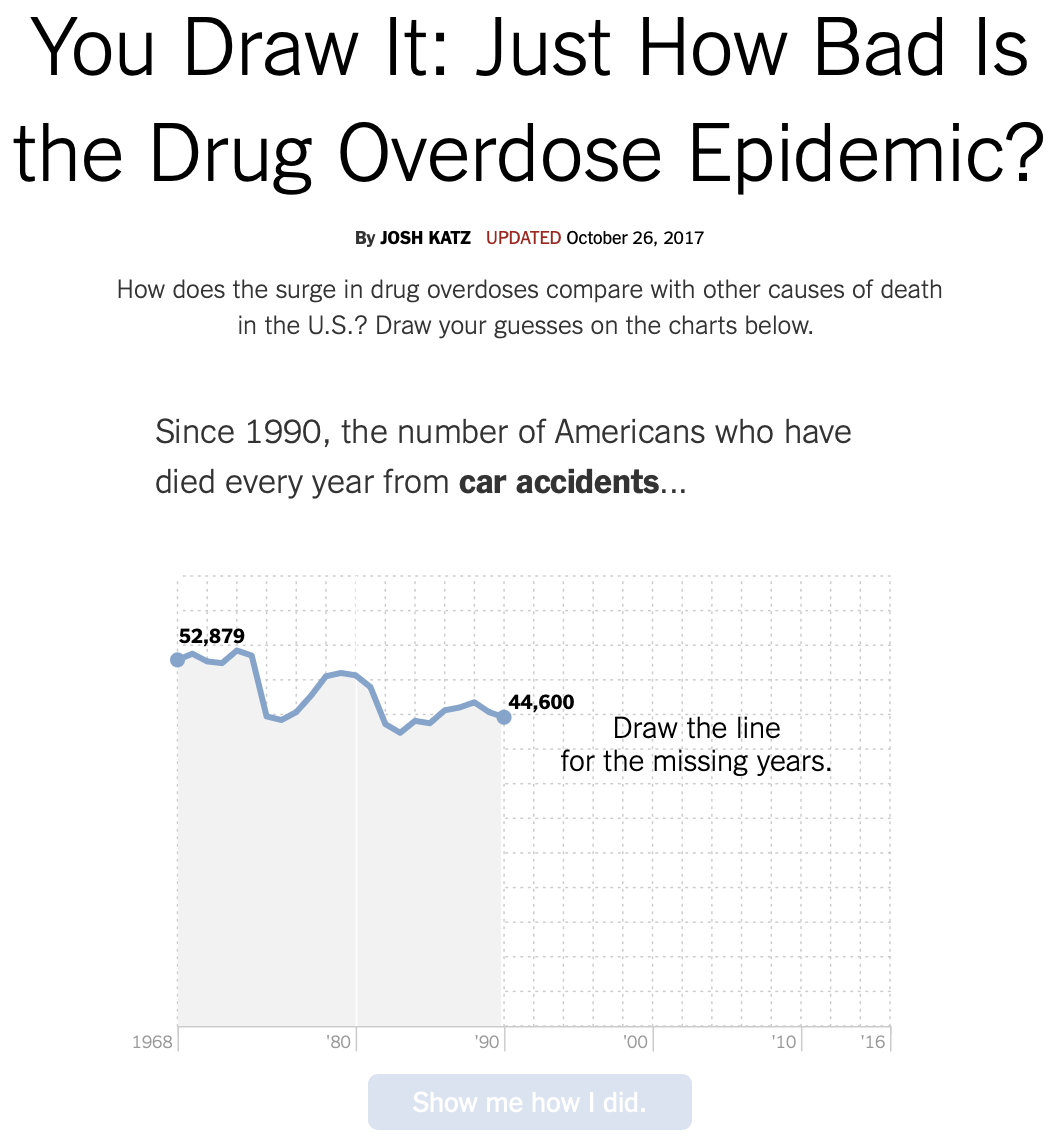}
    \caption{
  	``You Draw It" published in TheUpshot of New York Times \cite{YouDrawIt}.
  }
  \label{fig:NYT_you_draw_it}
\end{wrapfigure}

\IEEEPARstart{I}{magine} one sunny morning, having a freshly brewed cup of coffee by your side and you are ready to catch on up some news. A few headlines caught your eye. You read a couple of articles, scanning over texts and occasionally data visualizations illustrating data trends. The next article (an example shown in Figure \ref{fig:NYT_you_draw_it}) spurred your interest because it asked you to draw your guess before revealing the visual and textual content. 

When reading new articles we receive and evaluate new information. During this process individuals can fall anywhere along an ``elaboration continuum", i.e. they can think a lot, a moderate amount, or very little about the information \cite{wagner2011elaboration}. 
How do people absorb new information in order to inform their attitudes toward certain issues? Data visualization has played an increasingly important role in presenting new and evolving information to inform the public about data collected on high-stake issues (e.g. the drug overdose epidemic, Covid-19, the president's approval rating) \cite{YouDrawIt, HowUnpopularisJoeBiden}. Given the increased use of data visualizations in news articles \cite{FiveThirtyEight, Upshot}, is there anything visualization researchers and designers can do to create higher engagement and ``nudge" individuals to elaborate more? 

More elaboration has been theorized to lead to higher likelihood of attitude change in the Elaboration Likelihood Model of Persuasion (ELM) \cite{PettyELM}.
More specifically, ELM developed by Petty and Cacioppo as a theory to explain the process involved in attitude change and how these changes maintain themselves overtime \cite{PettyELM}, captured dual routes to attitude change. The ``central" and ``peripheral" routes to attitude change are distinguished by the degrees of elaborative thinking \cite{PettyELM_1999}.

While many studies have provided empirical evidence that existing attitudes are resistant to change \cite{Markant_2023, Heyer_20, Pandey_2014, Liem_2020}, there is a research gap in the field of data visualization on whether the attitude change or the lack thereof is based on an accurate internal representation of data visualizations individuals viewed. In addition, how is attitude change impacted by user engagement, as well as factors related to the individuals such as topic personal relevance in the context of data visualizations?
To address this gap, in this paper we investigate whether individuals update their attitudes after viewing a series of news articles related to a central theme, and the relationship between attitude change with recall of the data visualizations and engagement with the news articles. %their engagement with the news articles, as well as the relationship between recall and engagement with attitude change. 

Inspired by the ``You Draw it: Just How Bad is the Drug Overdose Epidemic?" article by the New York Times \cite{YouDrawIt} (Figure \ref{fig:NYT_you_draw_it}), we developed and incorporated the techniques--i.e. (1) visual elicitation by drawing a trend line and (2) contrasting narratives in our experiments. Contrasting narratives show drastic different trends compared to the main charts with accompanying textual descriptions. We aim to investigate the impact of elicitation and contrasting narratives on memory, engagement, and ultimately attitude change.

Our paper makes the following contributions: 
\revAliFinal{(1) Inspired by the ``You Draw it" article \cite{YouDrawIt}, we implement and compare elicitation methods --  including Draw trend (Figure \ref{fig:study_procedure} A) and Categorize trend (a simple drop-down with fixed-choice categorical answers) to elicit users' prior knowledge on temporal trends/predictions.} The Draw trend elicitation method produces a high-fidelity externalization of user's prior knowledge related to a temporal trend; data points and the overall trend were recorded by this elicitation method. To further ascertain whether any impact we observe is due to the elicitation step or the visual nature, we utilize the Categorize Trend elicitation method that asks users to choose a general characterization of the temporal trend they are asked to predict with a drop-down menu. 

(2) We offer empirical evidence on the impact of \revAliFinal{using such elicitation techniques in data communication} on attitude change, recall, and engagement by conducting a controlled experiment. We found overall participants showed significant attitude change regardless of elicitation conditions. Comparing across the elicitation conditions, our findings indicate that participants in the draw trend condition reported significantly higher engagement in terms of finding the article content surprising. In terms of recall error, while participants in the draw trend condition exhibited significantly lower error compared to the categorize trend condition, the difference was not significant compared to the control condition.

(3) A follow-up experiment that evaluates the addition of contrasting narratives to accent the main visualization in new articles. The followup study reused the three elicitation treatments in the study 1 and the only change was to add two contrasting narratives per article prior to the visualization shown in study 1. Findings on the impact of elicitation and on attitude change are consistent with study 1. Empirical evidence on contrasting narratives showed they improve engagement in terms of surprise and interest. Interestingly adding contrasting narratives also resulted in higher recall error across all elicitation conditions, possibly due to higher cognitive load required to recall information processed from both the main and contrasting charts.

We connect the dots from all of the above quantitative findings on the impact of elicitation and contrasting narratives in news articles with data visualization. Our discussion explores the relationship between attitude change, recall, and engagement informed by our studies.

% \IEEEPARstart{T}{his} file is intended to serve as a ``sample article file''
% for IEEE journal papers produced under \LaTeX\ using
% IEEEtran.cls version 1.8b and later. The most common elements are covered in the simplified and updated instructions in ``New\_IEEEtran\_how-to.pdf''. For less common elements you can refer back to the original ``IEEEtran\_HOWTO.pdf''. It is assumed that the reader has a basic working knowledge of \LaTeX. Those who are new to \LaTeX \ are encouraged to read Tobias Oetiker's ``The Not So Short Introduction to \LaTeX ,'' available at: \url{http://tug.ctan.org/info/lshort/english/lshort.pdf} which provides an overview of working with \LaTeX.

% \section{The Design, Intent, and \\ Limitations of the Templates}
% The templates are intended to {\bf{approximate the final look and page length of the articles/papers}}. {\bf{They are NOT intended to be the final produced work that is displayed in print or on IEEEXplore\textsuperscript{\textregistered}}}. They will help to give the authors an approximation of the number of pages that will be in the final version. The structure of the \LaTeX\ files, as designed, enable easy conversion to XML for the composition systems used by the IEEE. The XML files are used to produce the final print/IEEEXplore pdf and then converted to HTML for IEEEXplore.

% \section{Where to Get \LaTeX \ Help --- User Groups}
% The following online groups are helpful to beginning and experienced \LaTeX\ users. A search through their archives can provide many answers to common questions.
% \begin{list}{}{}
% \item{\url{http://www.latex-community.org/}} 
% \item{\url{https://tex.stackexchange.com/} }
% \end{list}

\section{Related Work}

\subsection{Effect of elicitation}
%Uncertainty ratings can improve the estimation of memory precision by several orders of magnitude https://onlinelibrary.wiley.com/doi/abs/10.1111/cgf.14556 VIBE: A Design Space for Visual Belief Elicitation in Data Journalism. 
A few studies in the field of visualization explored belief elicitation as a way to identify belief change and the relationship between belief elicitation and exploration behavior and discovery.
Karduni et al. evaluated a visual elicitation technique designed to elicit belief on a linear correlation and its uncertainty \cite{Karduni_beliefUpdating} and found that the ``line+cone" method produced comparable results to the MCMC-P method \cite{NIPS2007_MCMCP, SANBORN201063} that requires repeating the elicitation step many times until distribution convergence. The visualization elicitation technique can be applied before and after an intervention to capture belief change and the associated uncertainty. Mahajan et al. presented VIBE - a design space for visual belief elicitation in data journalism \cite{VIBE}. Many existing belief-driven visualizations from popular news media were discussed within the design space and trends and opportunities were highlighted.

Other studies evaluated the relationship between elicitation and exploration behavior with visual analytics systems. Through evaluating a tool designed to elicit users' beliefs and test their beliefs against data, Koonchanok et al. found that with belief elicitation participants were more likely to attend to discrepancies between their mental models and the data but less likely to engage in exploration, ultimately resulting in fewer discoveries \cite{koonchanok2021data}. In a follow-up study evaluating visual belief elicitation on scatterplots, Koonchanok et al. found that visual belief elicitation on correlation led to more accurate inferences and fewer false discoveries \cite{koonchanok2023visual}.

This paper adds to the current literature on evaluating the impact of elicitation on attitude change, recall, and engagement by implementing and designing two temporal elicitation methods, Draw trend and Categorize trend.

\subsection{Data visualization and belief and attitude change}

%Prior beliefs can lead to biased interpretations of statistical data extracted from visualizations. Xiong et al.\cite{xiong2022seeing} found that when a viewer has a strong prior belief that two variables are related, the viewer tends to overestimate the correlation, while underestimating the correlation if they held a strong belief that the two variables are unrelated. In another study, Xiong et al.\cite{Xiong2023} demonstrated that when exposed to contradictory trends, participants tend to weigh the positive trend more. Participants tend to favor the less sloped trend when both trends point in the same direction.
Xiong et al. discovered that prior beliefs can bias interpretations of data visualizations, leading viewers to overestimate correlations between variables they believe are related, and underestimate them when they think variables are unrelated \cite{xiong2022seeing} . Furthermore, they found that participants exposed to contradictory trends favor the positive one, and prefer the less steep trend when both point in the same direction \cite{Xiong2023}.

People can be reluctant to update their prior beliefs, especially when given a reason not to.  Research suggests that people are more conservative in updating their beliefs when the data presented in scatterplots are incongruent with their prior beliefs \cite{Markant_2023, xiong2022seeing}. Xiong et al. discovered that when the variables are ``belief-triggering," there is minimal change between the prior and post beliefs \cite{xiong2022seeing}.
Kim et al. found that participants were less likely to update their belief when other people’s expectations aligned with their own initial
expectations but not with the data \cite{Hullman2018DataThrough}.

Few studies have been conducted to model the effect of prior belief on visualization interpretation. %Kim et al. \cite{kim2019bayesian} used a Bayesian Cognition model to assess how people's beliefs change in response to visualizations. They discovered that, while people update their beliefs rationally on average, they frequently deviate from rational updates, especially when presented with a large sample size. 
Kim et al. \cite{kim2019bayesian} utilized a Bayesian Cognition model to study belief changes in response to visualizations, finding that although people generally update their beliefs rationally, they often deviate from rationality, particularly with large sample sizes.
Similarly, Karduni et al. \cite{Karduni_beliefUpdating} created a Bayesian Cognition model to predict belief updating with uncertainty visualizations. The model revealed no significant belief changes across various uncertainty representations, but indicated a shift in participants' confidence in their judgments.
%Karduni et al. developed a Bayesian Cognition approach to evaluate and predict belief updating with uncertainty visualizations . Their model showed no significant belief update across different uncertainty representation treatments but a change in participants' confidence about their judgements. 
Gupta et al. \cite{gupta2021belief} investigated how beliefs update with new data and the tendency to revert to original beliefs over time. They identified patterns of belief decay (reverting to initial beliefs) and belief persistence (maintaining updated beliefs) in participants.

\revDougFinal{Different from prior work on belief/attitude update with visual and uncertainty representations, the current studies focus on evaluating the impact of elicitation techniques on attitude change.}

\subsection{Data visualization and recall}

A few studies have focused on recalling from data visualization and explored potential systematic biases in recall.

An empirical study revealed systematic memory bias in participants replicating previously seen bars or dots, with overestimation of higher values and underestimation of lower ones \cite{mccoleman2020no}. Additionally, when a reference was present, responses around the 50\% mark showed underestimation below and overestimation above this threshold.
Similarly, Ceja et al. found that the aspect ratio of the bar marks bias how the position encodings are recalled \cite{Ceja_2021}. 

Kim et al. \cite{Hullman2018DataThrough} studied the effect of presenting social information as others' expectations instead of data alone. They found that participants remember data more accurately when it matches their expectations, and tend to maintain their own expectations and distrust the data when it aligns with others' but differs from the actual data. Kim et al. also evaluated how visually eliciting forms of prior knowledge and presenting feedback on the gap between prior knowledge and the observed data impact recall and comprehension \cite{Yea-Seul2017}. They found that participants who are prompted to reflect on their prior knowledge by predicting and self-explaining data outperform a control group in recall and comprehension when knowing little about the datasets.

Our studies build on findings from \cite{Yea-Seul2017} but have many distinctions. Primarily, we explore the impact of elicitation on attitude change which was not studied in \cite{Yea-Seul2017}. Moreover, the visual elicitation in \cite{Yea-Seul2017} was to elicit a few (discreet) missing data points while we contributed a visual elicitation technique on temporal trends. In addition to evaluating the impact of elicitation, our second study focuses on evaluating the impact of contrasting narratives on recall, engagement, and attitude change.

\subsection{Attitude change and Elaboration Likelihood Model}
The Elaboration Likelihood Model (ELM) \cite{PettyELM_1999, PettyELM} theorizes how factors like expertise, message content, recipient mood, and context affect attitude change. It outlines a dual-route process—central and peripheral—determining the degree of elaborative thinking \cite{PettyELM_1999}. The central route leads to lasting attitude changes resistant to counter-persuasion, whereas the peripheral route results in temporary changes. ELM also identifies three processes where no attitude change occurs despite some elaborative thinking.
%The elaboration likelihood model of persuasion (ELM)  was formulated as a theory about how multiple factors, including knowledge sources (e.g., expertise), message (e.g., number of arguments), recipient (e.g., mood), and contextual (e.g., distraction) variables have an impact on attitude change. The model was schematically depicted in a decision tree-like presentation and articulates a dual route with multiple processes in which variables can impact judgements. The dual routes in ELM refers to the ``central" and ``peripheral" routes to attitude changes based on different degrees of elaborative thinking . The central route in the ELM leads to more enduring attitude change that is more resistant to counter-persuasion, while the peripheral route leads to a relatively temporary attitude change. ELM also presented three processes that result in no attitude change despite a certain amount of elaborative thinking taking place. 
We use the ELM model as a schema to inform our hypothesis development and as a way to understand and discuss our experiment results.

\section{Hypothesis Development}
Two aspects that made the New York Times ``You Draw It" article on the topic of drug overdose epidemic particularly interesting to us as visualization researchers are the use of \textit{visual elicitation} and \textit{\textcolor{revision}{contrasting narratives}} to accent the main data visualization and message focused on rising deaths from drug overdoses over time. 

\textcolor{revision}{Instead of just viewing the chart, elicitation prompts users to first externalize their belief/expectation on the issue. Prior research showed displaying the real data trend against one's own expectations leads to more attention to the discrepancies \cite{Yea-Seul2017}. Moreover, Kirsh discussed seven ways that people can \revAliFinal{``think more powerfully" with external representations }
%enhance people's cognitive power
\cite{kirsh_2010}. Since users may produce such external representations of their belief/expectations with different resolutions, we designed a non-visual elicitation that simply ask people to categorically characterize the trend compared to the draw-trend visual elicitation method used in the New York Times article \cite{YouDrawIt}. %Therefore, we hypothesize that visual elicitation may lead to better recall and engagement. 
Similarly, contrast may lead to better recall and engagement since they provide a ``surprise" factor that accent the severity of the main issue the article is trying to convey \cite{brod2018generating,fazio2009surprising}.} Connecting to the Elaboration Likelihood Model \cite{PettyELM}, we contend that elicitation and \textcolor{revision2}{contrasting narratives} may lead to higher amounts of elaborative thinking \textcolor{revision}{compared to passive viewing} and ultimately result in a higher chance of attitude change. 
%While many prior experiments evaluated belief and attitude change with data visualizations, \textcolor{revision}{few have analyzed whether attitude change is related to accurate memory of the data trends shown and individuals' engagement. In other words, could a user update their attitude without being able to accurately recall the data trends?} %the impact of elicitation and \textcolor{revision2}{contrasting narratives} on . 

\subsection{Hypothesis formulation}
Three main hypotheses were formulated:

%\textbf{H1-Attitude change.} Our first hypothesis focuses on whether individuals exhibit attitude change after viewing a series of three news articles with data visualization related to the drug overdose epidemic. \textcolor{revision}{We hypothesize that both belief elicitation and \textcolor{revision2}{contrasting narratives} lead to higher likelihood of attitude change compared to when these two techniques were not present in the news articles.}%We will also evaluate the impact of belief elicitation and \textcolor{revision2}{contrasting narratives} on attitude change. 

\textcolor{revision2}{\textbf{H1-Impact of elicitation on recall and engagement.} }
%(drawing a trend and categorizing a trend) 
Belief elicitation increases recall and engagement with news articles that contain data visualizations compared to the condition with no elicitation. 

Expanding on H1, we expect a difference in the impact of the two forms of elicitation (draw a trend vs. categorize a trend) on recall and engagement. Compared to categorizing a trend as a slight or sharp increase, individuals in the draw a trend condition are asked to draw a line across multiple years, producing data points for each year as well as a trend line. Thus individuals in the ``draw a trend" condition generate a higher-fidelity representation of their guess/expectation.

\textcolor{revision2}{\textbf{H2-Impact of contrasting narratives on recall and engagement.} } Adding contrasting narratives to accent the data visualization on the article's focal issue increases recall and engagement with news articles. %In addition, adding contrasting narratives increases the likelihood of attitude update.

\textcolor{revision2}{\textbf{H3-Impact of elicitation and contrasting narratives on attitude change.} }While prior research has demonstrated people's attitudes are notoriously difficult to change, we hypothesize belief elicitation and contrasting narratives increase the likelihood of attitude update on the main issue presented in the news articles.
%we'd like to explore the effect of the combination of belief elicitation and contrast. We expect the combination to be more powerful than belief elicitation or contrast alone.

\begin{figure}[t]% specify a combination of t, b, p, or h for top, bottom, on its own page, or here
  \centering
  \includegraphics[width=\columnwidth]{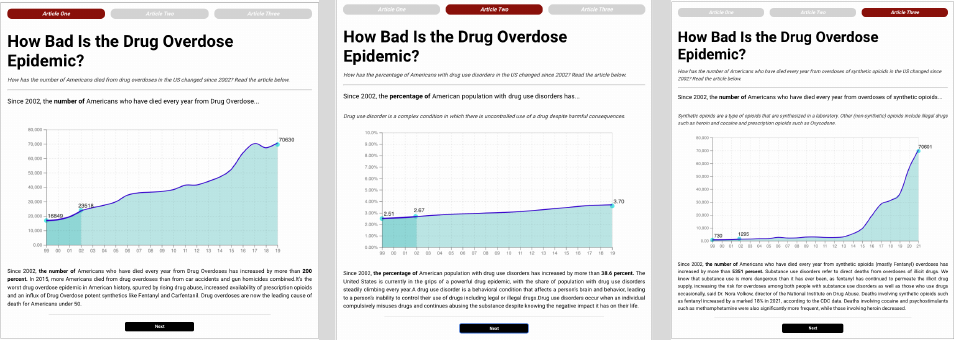}
  \caption{%
  	Three articles participants read during Study 1 regardless of elicitation conditions.
  }
  \label{fig:study1_stimuli}
\end{figure}

\section{Study design}

\subsection{Datasets and stimuli}

The New York Times ``You Draw It" article focuses on rising deaths from the drug overdose epidemic in the US \cite{YouDrawIt}. For our studies, we kept the theme of the drug overdose epidemic and generated two additional articles, with one reporting the percentage of the American population with drug use disorders and the other reporting deaths from a particular type of illicit drug (synthetic opioids). \revAli{We believe due to the current polarized conversations around the opioids epidemic, this is an appropriate theme to address and learn about how visualizations and narrative might help with attitude change around polarizing topics.}

\textcolor{revision}{\textbf{Study 1 -} }Yearly data for drug overdoses and the percentage of the American population with drug use disorders were collected from \textit{Our World in Data }\cite{DrugOverdose}. 
Data for the number of Americans who have died every year from overdoses of synthetic opioids was collected from \cite{DrugOverdoseDeathRates}. 
The stimuli participants saw regardless of elicitation conditions are shown in Figure \ref{fig:study1_stimuli}.

\begin{wrapfigure}{l!}{0.18\textwidth} %this figure will be at the right
    \centering
    \includegraphics[width=0.18\textwidth]{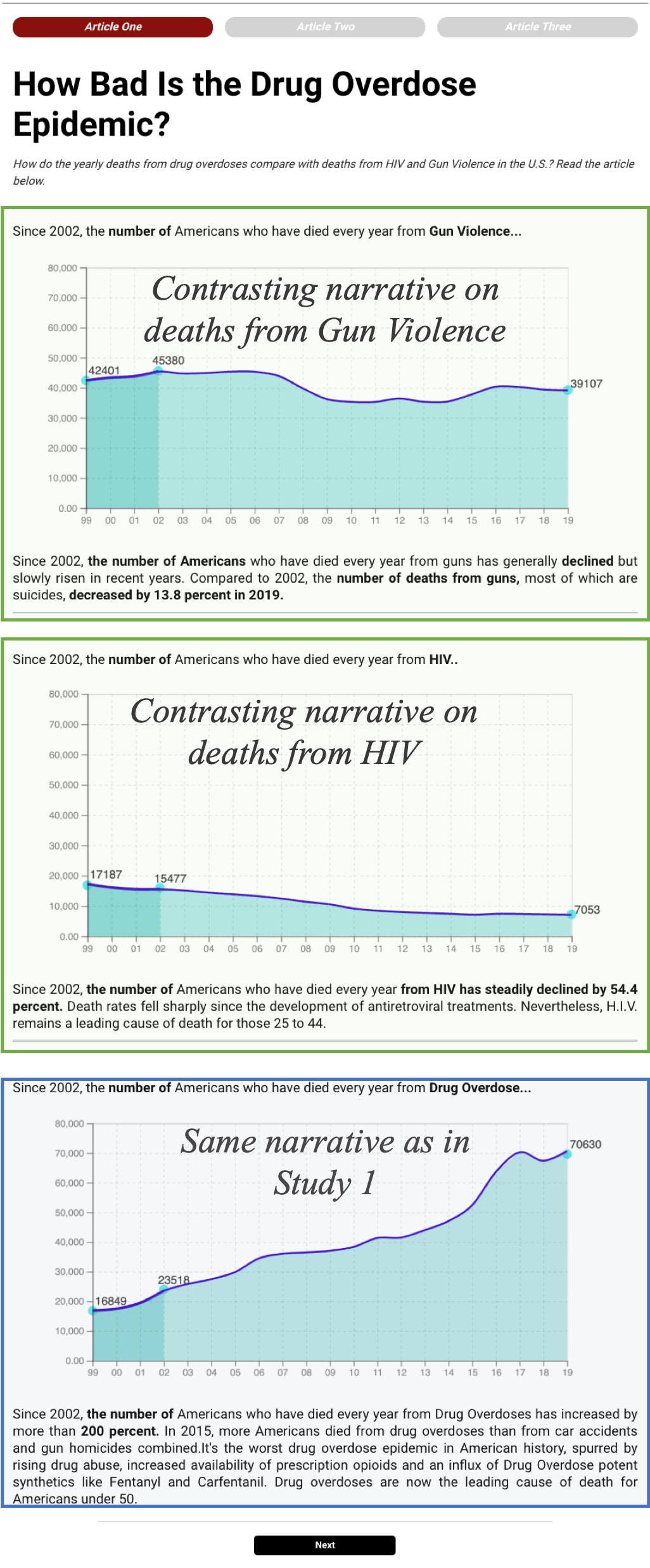}
    \caption{
  	Contrasting narratives on \textit{gun violence} and \textit{HIV} deaths added to contrast the sharp rise in drug overdose deaths.
  }
  \label{fig:contrast_charts}
\end{wrapfigure}

\begin{figure*}[b]% specify a combination of t, b, p, or h for top, bottom, on its own page, or here
  \centering
  \includegraphics[width=.9\linewidth]{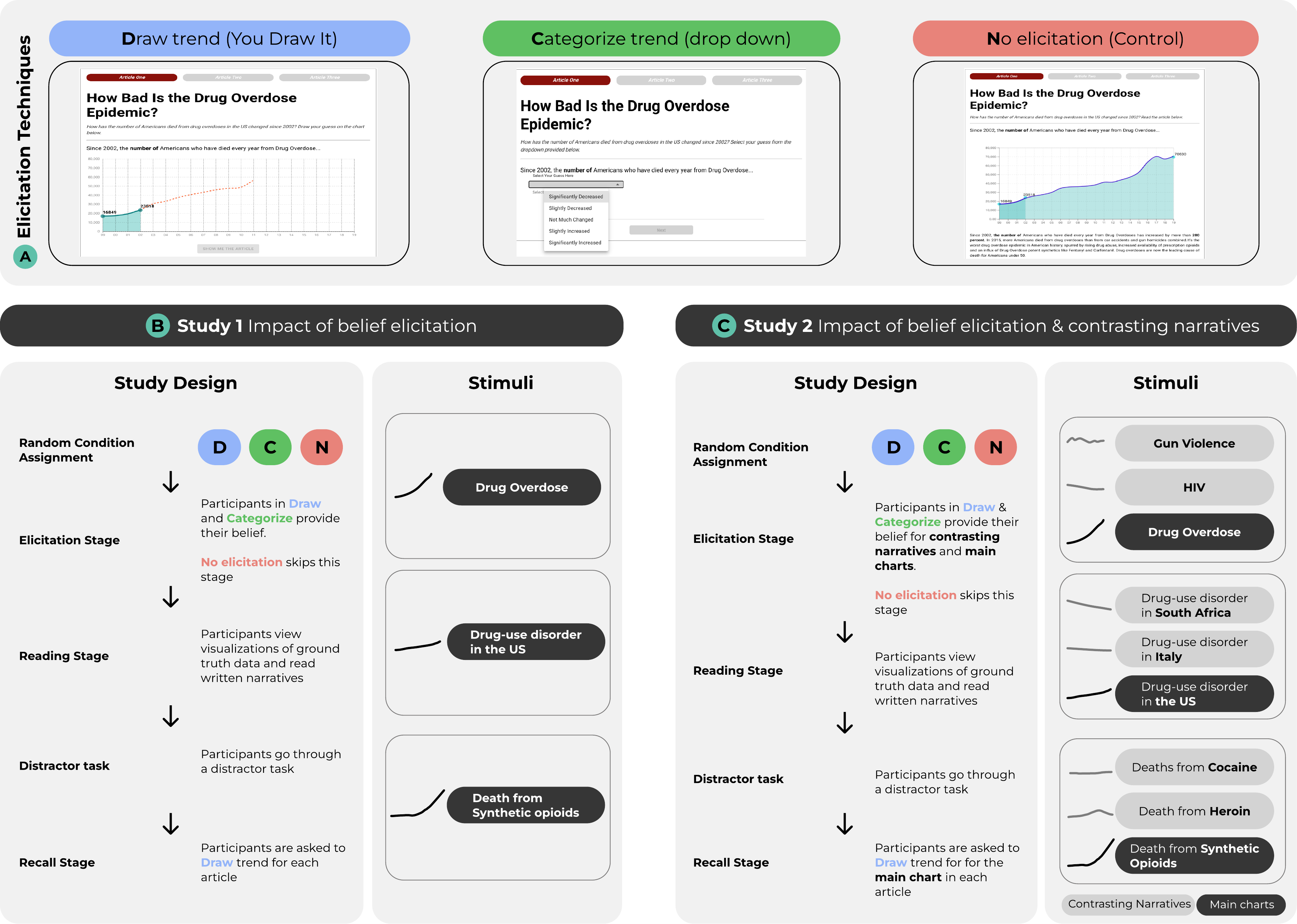}
  \caption{%
  	Study design and procedure. Study 1 focuses on evaluating the impact of belief elicitation while study 2 added \textcolor{revision2}{contrasting narratives} to evaluate its impact on attitude change, recall, and engagement.
  }
  \label{fig:study_procedure}
\end{figure*}

\textcolor{revision}{\textbf{Study 2 -} Keeping the same stimuli as in study 1,} we added two contrasting trends to each of the articles as detailed in the bottom right portion of Figure \ref{fig:study_procedure}. \textcolor{revision2}{We refer to these contrasting trends with brief textual descriptions of the trends as \textit{contrasting narratives}.} 
For example, visualizations of deaths from Gun violence and HIV with accompanying text descriptions of the trends (top two cells in Figure \ref{fig:contrast_charts}) were added to article 1 before participants saw the visualization on deaths from drug overdoses (bottom cell in Figure \ref{fig:contrast_charts}). 
The purpose of adding the \textcolor{revision2}{contrasting narratives} is to accent the last chart in each article related to the drug overdose epidemic in the US. \textcolor{revision}{The last chart in each article was the same chart from Study 1.} An example of article 1 (in the control condition) is shown in Figure \ref{fig:contrast_charts}. %
  \\%

\subsection{Experiment conditions design rationale}

In our experiments, the ``Draw trend" condition refers to asking participants to complete a timeline by dragging the cursor in the chart area ( example shown in Figure \ref{fig:NYT_you_draw_it}). The participants can adjust any individual  data point to shape the line to their desired trend. This technique is similar to the one used in the drug overdose epidemic NYT article \cite{YouDrawIt}.  
To ascertain whether the impact of belief elicitation is due to the asking participants to guess the trend first before seeing the data or the visual nature of drawing a trend, we designed the ``Categorize trend" condition that asks the participants to select a general characterization of the temporal trend (e.g., slight increase, mostly flat, slight decrease, etc.) from a dropdown menu (Figure \ref{fig:study_procedure}A). The control condition does not present any form of elicitation; the visualization and accompanying textual description are shown to individuals, similar to the most common way we view news articles.
  
In summary, we designed three conditions for study 1, which focuses on evaluating the impact of belief elicitation: \textbf{\textcolor{cyan}{Draw trend}}, \textbf{\textcolor{ForestGreen}{Categorize trend}}, and \textbf{\textcolor{Mahogany}{No elicitation}}.

Building on study 1, \textcolor{revision2}{we further} explore the effect of adding \textcolor{revision2}{contrasting narratives} that accent the main visualization. The three elicitation conditions remain the same in Study 2, and the only difference is adding two \textcolor{revision2}{contrasting narratives} prior to the focal visualization present in Study 1 in each article. The design of Study 2 allows us to explore the potential interaction between elicitation and contrasting narratives. 

\subsection{Study procedure}
\textcolor{revision}{\textbf{Study 1} - Detailed procedure of study 1 is illustrated in Figure \ref{fig:study_procedure}B.}
The goal of Study 1 was to evaluate the impact of elicitation on memory, engagement, and attitude change. 
\textcolor{revision}{\textbf{Study 2} - Detailed procedure study 2 is illustrated in Figure \ref{fig:study_procedure}C.}
The goal of Study 2 was \linebreak[2]to evaluate the effect of \textcolor{revision2}{contrasting narratives} on memory, engagement, and attitude change, and potential interactions between elicitation and contrasting narratives.
%Both studies employed a mixed design with a between-subjects manipulation of the elicitation treatment and a within-subject manipulation of the articles presented to participants(all participants saw three articles in the same order). 
\revDougFinal{Both studies employed a mixed design structure with 2 fixed and 1 random factors: (F1) elicitation technique (between-subjects) with 3 levels : none [control], draw trend, categorize trend, (F2) article (within-subjects) with 3 levels: drug-overdose, drug-use-disorder, Opioid-death, and (R1) participant levels. Participants were nested within elicitation conditions, and questions were fully crossed with condition. Thus, each participant was randomly assigned to one condition, in which they completed the task with all three articles (in the same order). }
The pre-registration of study 1 and 2 can be found here \footnote{\RaggedRight \small Study 1: \url{https://bit.ly/41l856z}, Study 2: \url{https://bit.ly/47Y061V}}

\noindent \textcolor{revision2}{In both studies, participants were instructed to engage in two distractor tasks between the reading phase and the recall phase. Initially, the participants responded to a series of inquiries, after which they engaged with a news article related to the 2023 Super Bowl \footnote{\RaggedRight \small \href{https://sports.yahoo.com/super-bowl-2023-poor-turf-was-issue-for-players-its-the-worst-field-i-ever-played-on-055124350.html}http://bit.ly/3RFsn7U}. Subsequently, they offered a one-sentence summary of the article, before proceeding to the recall phase. }

\noindent \revMilad{Topic involvement is elicited once per participant prior to seeing any articles using the 4-item questionnaire developed by Liao et al. \cite{TopicInvolvement}.} \revDougFinal{Topic involvement measures personal relevance and motivation to learn about controversial topics \cite{TopicInvolvement} and was shown in prior research to impact belief and attitude change \cite{Markant_2023}. We use the measure of topic involvement when analyzing attitude change in our model and interpret results in the context of this factor.}

\noindent We collected data from 288 participants in Study 1 and 307 participants in Study 2. \textcolor{revision}{There was no overlap in participants between the two studies. Participants demographics from both studies share similar distributions.} Details on the participants demographics and exclusion of data can be found in section \ref{sec:participants}.

\subsection{Dependent variables}
\textcolor{revision}{Studies 1 and 2 shared the same sets of dependent variables.} %In this section we introduce the dependent variables we measured in order to evaluate the impact of elicitation \revDoug{method} and \textcolor{revision2}{contrasting narratives}. 
Recall accuracy and article engagement are measured at the article-level, resulting in a set of scores for each of the three news articles.
Attitude change is measured at the participant-level; i.e. one set of scores from three attitude questions per participant. 

\subsubsection{Recall error}

We calculated the Root Mean Square Error (RMSE) between the participant's recall response and the actual data for each chart to determine the overall recall accuracy for each user. RMSE is a widely used measure of error on numerical predictions \cite{RMSE_use}. As shown in Equation \ref{rmse}, $\hat{y}$ is the recalled value by a participant for each year, $y$ is the \textcolor{revision2}{ground-truth} value for that year in the dataset and $n$ is the total number of years in each response:

\begin{equation} \label{rmse} 
RMSE_{overall} = \sqrt{\sum{(\hat{y}-y)^2}/n}
\end{equation}

% DM: Commenting these out unless we end up using them in results

% \textbf{Overall signed recall error} To measure the overall error between each response and the true trend, we calculated the Signed Mean Recall Error (SMRE) by subtracting the mean of responses ($\bar{\hat{y}}$) from the mean of true trend ($\bar{y}$ ) for each chart as shown in Equation \ref{smre}.  

% \begin{equation} \label{smre} 
% \bar E_{overall} = (\bar{\hat{y}} - \bar{y} ) 
% \end{equation}

% \textbf{Yearly recall accuracy} We calculated a point-to-point error for each response per article to better understand the error variation across each participant's drawing. As shown in Equation \ref{sma}, the Signed Recall Accuracy (SRA) is the difference between the participant's response and the true value for that article for the year $i$.

% \begin{equation} \label{sma} 
% \bar E_{i} = (\hat{y}_i - \bar{y}_i ) 
% \end{equation}

\subsubsection{Article engagement}
We measured participants' engagement with the news articles as a proxy for the amount of elaborative thinking that occurs \revDoug{in response to each article.}
% while they performed the experiment tasks. 
We evaluate \textcolor{revision2}{three components of} user engagement (i.e. surprise, recommend, interest) \textcolor{revision2}{using} three questions adapted from the User Engagement Scale (UES) \cite{OBRIEN201828}. UES was designed to measure user engagement as a quality of user experience characterized by the depth of an user's investment when interacting \textcolor{black}{with} a digital system \cite{O'Brien_2016}. \textcolor{revision2}{The entire UES measures many dimensions including perceived usability and aesthetic appeal. Since not all dimensions apply to our evaluation of news article engagement, we selected a subset of questions to evaluate the following three components of user engagement:}

\textbf{Surprise:} \textit{``The content of this article is surprising to me."}

\textbf{Recommend:} \textit{``I would recommend this article to my family and friends."}

\textbf{Interest:} \textit{``I felt interested in this article."}

Responses to each question were on a 5-point scale with the responses ``Not at all", ``A little", ``Moderately", ``A lot", and  ``Extremely."

\subsubsection{Attitude change}
We designed three questions to measure participants' attitudes regarding the drug overdose epidemic. Responses to these three questions were made on 5-point scale.
The three questions and corresponding response scales are: 

\textit{Q1. To what extent do you think the current rate of drug overdoses in the US is a problem?} Responses to this question ranged from ``Extremely serious problem" to ``Not at all a problem".

\textit{Q2. Should the US government make combating drug abuse and overdoses a priority, i.e. by allocating tax dollars to treatment and prevention programs?} Responses to this question ranged from ``Not a priority" to ``High priority". 

\textit{Q3.  What is your opinion on drug legalization and decriminalization in the US?} Responses to this question ranged from ``Strongly Oppose" to ``Strongly Favor." 

The first two attitude questions are designed to evaluate the perceived seriousness and whether the participant is willing to direct more resources to address the problem. The first two questions are one directional; we expect responses to indicate the drug overdose epidemic to be more or less serious. 
With the last question, we were interested in how people think of drug legalization when provided more information on the current drug overdose epidemic in the US.

\subsection{Overview of analysis approach}

All analyses were conducted in R version 4.2.3. %~\cite{RManual}
Data and analysis code are available at \footnote{the Supplemental Materials is available here: \url{https://bit.ly/3GJoAjE}}.
We performed Bayesian statistical analyses using \textit{brms} \cite{burkner2017brms}. To evaluate the fitted Bayesian models (e.g. baseline vs. models with additional fixed effects) we used LOO-PSIS \cite{vehtari2017practical}, a robust method for comparing the performance of Bayesian models in terms of expected predictive accuracy.
\revDougFinal{As described in our preregistration, as a general practice we compared two models for each dependent measure. The baseline model included treatment (Control, Categorize, Draw) and article (Drug Overdoses, Drug-use Disorder, and Synthetic Opioids) as fixed effects and a random intercept for participants (\texttt{y $\sim$ treatment + article + (1|PID)}).}\footnote{\revDougFinal{In our preregistration we planned to include article (stimulus ID) as a random effect. However, given the small number of stimulus categories (3 articles) we chose to include it as a fixed effect.}} 
We compared that to an alternative model that also included topic involvement and prior attitude as additional fixed effects
\revDougFinal{(\texttt{y $\sim$ treatment + article + 
topicInvolvement + preAttitude + (1|PID)}), based on the expectation that participants' existing views on the topic might influence their engagement in the task and responses to the articles.}
\revDougFinal{In all cases, model comparisons based on the LOO-PSIS criterion were either ambivalent or favored the more complex model, so for simplicity we only report the results for the alternative model.}
\revDougFinal{In these and all subsequent models, treatment coding was used for categorical predictors with the Control condition and Drug Overdoses article serving as reference levels. 
Full results for estimated parameters of each model are provided in the Supplementary Material.}

In our results we report posterior medians and 95\% highest density intervals (HDIs) for estimated parameters and contrasts. 
For focal effects of interest we also report the \textit{probability of direction (PD)} which ranges from 50--100\% and \revDoug{is the proportion of the posterior distribution that is above zero (when the posterior median is positive) or below zero (when the posterior median is negative) \cite{makowski_indices_2019}. PD thus indicates the overall probability that an estimated quantity is in the same direction as indicated by the posterior median.} 
For ease of interpretation we highlight those effects for which the 95\% HDI does not include zero (or another null value). 
However, we note that under the Bayesian estimation approach the posterior distribution can be directly interpreted as the relative credibility of different parameter values or effect sizes, and a 95\% HDI that overlaps with zero can still provide (weaker) evidence for non-zero effects \cite{kruschke_bayesian_2018}.

\section{\textcolor{black}{Results on H1}: The impact of elicitation}

\label{sec:elicitation_impact}

\subsection{The impact of belief elicitation on recall error}

\begin{figure}[tb]% specify a combination of t, b, p, or h for top, bottom, on its own page, or here
  \centering % avoid the use of \begin{center}...\end{center} and use \centering instead (more compact)
  \includegraphics[width=0.8\columnwidth]{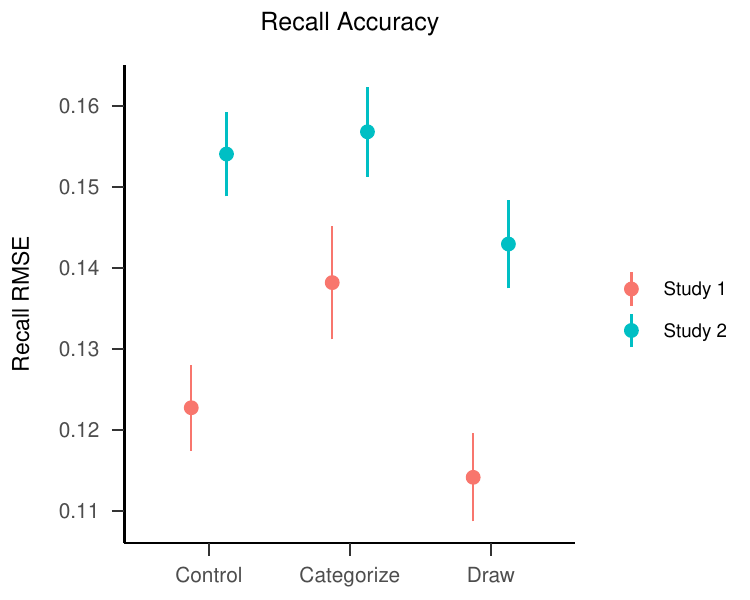}
  \caption{\revDougFinal{Recall accuracy (RMSE) by condition for Study 1 and Study 2. Error bars indicate standard error of the mean.}}
  \label{fig:recall-rmse}
\end{figure}

This set of analyses focused on participants' accuracy at the recall stage of the task, when they were asked to recreate the trends for the three articles shown earlier. 
We normalized the RMSE values by dividing each data point by the maximum of that article. As a result, recall RMSE values can be interpreted as errors \revDougFinal{as a percentage of the displayed height of the chart}.
Figure \ref{fig:recall} (top) shows the distribution of participants' responses alongside the true trend (black line), and the year-by-year absolute error for each treatment (bottom). 
\revDougFinal{Figure \ref{fig:recall-rmse} shows the overall RMSE aggregated by treatment.}
For each study we modeled overall recall error (RMSE) using multilevel GLMs with a log-normal link function.

\textbf{Study 1-Elicitation}: 
\revDougFinal{The results from pairwise contrasts between treatments} are shown in Table \ref{tab:studiesRMSE}a.  
Overall we did not observe credible difference when comparing either belief elicitation method (Draw or Categorize) to the control condition with no elicitation. However, when comparing two methods of elicitation, our results indicated that recall RMSE was lower in the Draw trend elicitation compared to the categorize trend elicitation condition (Draw -- Categorize = $-.02$, 95\% HDI [-.04, -.003]), with $PD = 98\%$ probability of a negative effect.
\revAliFinal{This result suggests that participants in the Draw condition were, on average, likely to be more accurate in recreating the trend than participants in the Categorize condition.}

\begin{table}[htb]
\centering
\footnotesize
\begin{tabular}{@{}ccll@{}}
\toprule
\multicolumn{4}{c}{\textbf{Pairwise Recall RMSE}}                            \\ \midrule
\rowcolor[HTML]{EFEFEF} 
\multicolumn{1}{l}{\cellcolor[HTML]{EFEFEF}\textbf{Treatment Pairs}} &
  \textbf{Recall RMSE} &
  \multicolumn{1}{c}{\cellcolor[HTML]{EFEFEF}\textbf{95\% HDI}} &
  \multicolumn{1}{c}{\cellcolor[HTML]{EFEFEF}\textbf{PD}} \\ \midrule
\multicolumn{4}{c}{\textbf{a: Study 1}}                            \\ \midrule

\textbf{Draw - Categorize}    & \textbf{-0.02}                     & \textbf{{[}-0.04 , -0.003{]}} & \textbf{98\%}   \\
Categorize - Control & 0.01                      & {[}-0.01, 0.03{]}    & 87\%   \\
Draw - Control       & -0.01                     & {[}-0.03, 0.005{]}   & 83\%   \\ \midrule
\multicolumn{4}{c}{\textbf{b: Study 2}}                                     \\ \midrule
\textbf{Draw - Categorize}    & \multicolumn{1}{c}{\textbf{-0.02}} & \textbf{{[}-0.04, -0.001{]}}  & \textbf{98.6\%} \\
Categorize - Control & \multicolumn{1}{c}{0.004}  & {[}-0.01, 0.02{]}    & 71\%   \\
Draw - Control       & \multicolumn{1}{c}{-0.01} & {[}-0.03, 0.003{]}    & 93\%  \\
\bottomrule\\
\end{tabular}%

\caption{Pairwise Recall RMSE Difference (including posterior median, 95\% High Density Intervals, and Probability of Direction) for Study 1 and 2.}
\label{tab:studiesRMSE}
\end{table}

% Revised figure

% Old figure
% \begin{figure}[tb]% specify a combination of t, b, p, or h for top, bottom, on its own page, or here
%   \centering % avoid the use of \begin{center}...\end{center} and use \centering instead (more compact)
%   \includegraphics[width=\columnwidth]{figs/model_recall_contrasts.pdf}
%   \caption{Contrasts between treatments for model of recall RMSE. \revDoug{Error bars indicate 95\% HDIs.}}
%   \label{fig:recall-rmse-contrasts}
% \end{figure}

\textbf{Study 2-Elicitation+Contrasts}: 
\textcolor{revision}{The elicitation results obtained from Study 2 were consistent with those of Study 1 (Table \ref{tab:studiesRMSE}b).} %\revAliFinal{%Recall error was lower in the Draw elicitation condition compared to the Categorize trend elicitation condition, while there was weaker evidence that Draw elicitation RMSE was lower than the control condition and no difference between the control and categorize trend conditions.
%Participants in the Draw elicitation condition were on average more accurate than the Categorize trend condition, while there was weaker evidence that Draw elicitation RMSE was lower than the control condition and no difference between the control and categorize trend conditions.}
\revAliFinal{Interestingly, the results for Studies 1 and 2 revealed that participants in the Draw elicitation condition were on average more accurate than the Categorize trend condition. However, recall errors from these two conditions were not significantly different than the Control condition. \revAliFinal{Overall, the magnitude of the effect on recall remains a question to be explored in future studies.} Further discussions of these findings are in section \ref{sec:discussion_elicitation}.}

\subsection{The impact of belief elicitation on engagement}
\label{sec:elicitation_and_engagement}

\begin{figure*}[tb]% specify a combination of t, b, p, or h for top, bottom, on its own page, or here
  \centering % avoid the use of \begin{center}...\end{center} and use \centering instead (more compact)
  \includegraphics[width=0.9\textwidth]{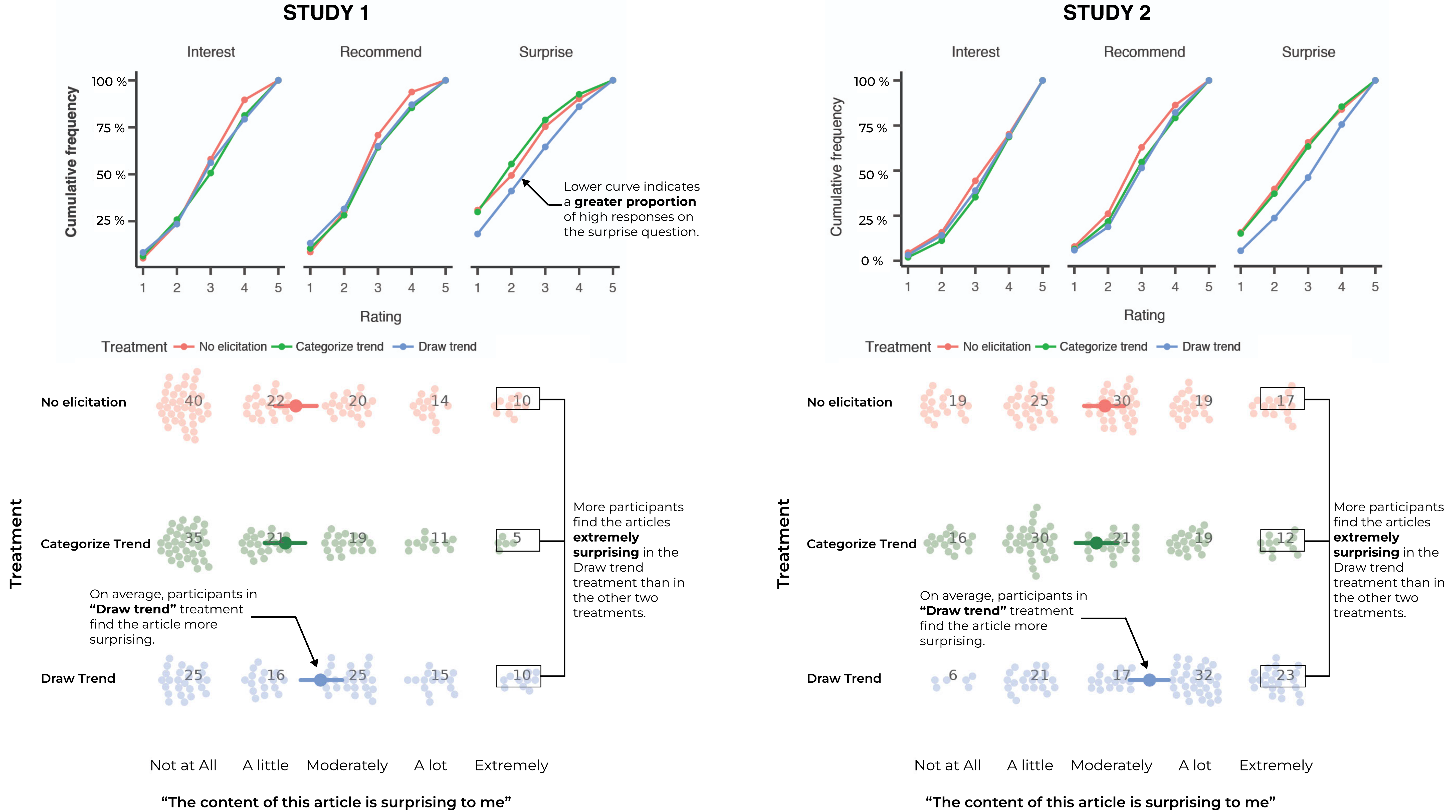}
  \caption{Top: Cumulative frequency of responses to the interest, recommend, and surprise questions for Study 1 \& 2. Bottom: distribution of individual responses across treatments for the ``surprise" question.
  }
  \label{fig:study1-engagement}
\end{figure*}

% \begin{figure}[tb]% specify a combination of t, b, p, or h for top, bottom, on its own page, or here
%   \centering % avoid the use of \begin{center}...\end{center} and use \centering instead (more compact)
%   \includegraphics[width=\columnwidth]{figs/study1/Surprise-01.pdf}
%   \caption{Top: Cumulative frequency of responses to the interest, recommend, and surprise questions for Study 1. Bottom: distribution of individual responses across treatments for the ``surprise" question.
%   }
%   \label{fig:study1-engagement}
% \end{figure}

Next we examined whether elicitation has an impact on participants' responses to the post-recall engagement questions (interest, recommend, and surprise).
Responses to each question were on a 5-point scale.
We calculated the proportion of participants in each treatment group who gave each response.
Figure \ref{fig:study1-engagement} (top) shows the empirical CDF obtained by calculating the cumulative sum across the five rating categories.
Lower curves represent treatments with a greater proportion of high responses on a given question. 
The \revDougFinal{blue} curve (top of Fig. \ref{fig:study1-engagement}) is lower for both Study 1 and 2, indicating more participants in the Draw condition find the content of the article surprising. Individual responses to the engagement question on ``surprise" are plotted under the curves. We will unpack \revDougFinal{these results for perceived surprise in the next section.}

\revDougFinal{We used ordinal regression to separately model responses for each outcome (surprise, interest, recommend).
The models used a cumulative logit link function with separate threshold parameters fit for each rating category.} 
% \revDougFinal{As with the models of recall RMSE, we first fit a baseline model with fixed effects for treatment and article and random intercepts for participants (\texttt{y $\sim$ treatment + article + (1|PID)}, with separate models fit for each outcome $y \in \{\texttt{surprise, interest, recommend}\}$). We compared this with a second model that also included fixed effects for topic involvement and prior attitude (\texttt{y $\sim$ treatment + article + 
% topicInvolvement + preAttitude + (1|PID)})}.
% \revDougFinal{For all three questions, model comparisons based on PSIS-LOO were either ambivalent or favored the more complex model, so for simplicity we only report the results of the second model.}

\textbf{Study 1}: \revAliFinal{
The results indicated that %perceived surprise was higher 
participants in the Draw condition were on average likely to be more surprised compared to the Control and the Categorize conditions, while there was no apparent difference between the Categorize and Control conditions (Table \ref{tab:studiesSurprise}a).
} 
%This indicated that perceived surprise was higher in the Draw condition compared to the Control condition (Draw - Control = .44, 95\% HDI [.06, .83], $PD = 98\%$) and the Categorize condition (Draw -- Categorize = .59, 95\% HDI [.17, .96], $PD = 99.7\%$), while there was no difference between the Categorize and Control conditions (Categorize - Control = .15, 95\% HDI [-.21, .56], $PD = 78\%$).
Looking at the \revDougFinal{distribution of} responses to the surprise question, Fig \ref{fig:study1-engagement} bottom left, fewer participants in the Draw condition found the articles ``Not at all" surprising while more participants responded ``Moderately" surprising.
There were no effects for the perceived interest and recommend questions.

\begin{table}[htb]
\centering
\footnotesize
\begin{tabular}{@{}ccll@{}}
\toprule
\multicolumn{4}{c}{\textbf{Pairwise Perceived Surprise}}                            \\ \midrule
\rowcolor[HTML]{EFEFEF} 
\multicolumn{1}{l}{\cellcolor[HTML]{EFEFEF}\textbf{Treatment Pairs}} &
  \textbf{Surprise} &
  \multicolumn{1}{c}{\cellcolor[HTML]{EFEFEF}\textbf{95\% HDI}} &
  \multicolumn{1}{c}{\cellcolor[HTML]{EFEFEF}\textbf{PD}} \\ \midrule
\multicolumn{4}{c}{\textbf{a: Study 1}}                            \\ \midrule

\textbf{Draw - Categorize}    & \textbf{0.53}            & \textbf{{[}0.15 , 0.99{]}} & \textbf{99.7}\%   \\
Categorize - Control          & -0.14                     & {[}-0.52, 0.23{]}    & 78.0\%   \\
\textbf{Draw - Control}       & \textbf{0.38}            & \textbf{{[}0.03, 0.77{]}}   & \textbf{98.0}\%   \\ \midrule
\multicolumn{4}{c}{\textbf{b: Study 2}}                                     \\ \midrule
\textbf{Draw - Categorize}    & \multicolumn{1}{c}{\textbf{0.53}} & \textbf{{[}0.15, .93{]}}  & \textbf{100.0}\% \\
Categorize - Control          & \multicolumn{1}{c}{-0.14}  & {[}-0.52, 0.23{]}    & 57.5\%   \\
\textbf{Draw - Control}       & \multicolumn{1}{c}{\textbf{0.38}} & \textbf{{[}0.02, 0.77{]}}    & \textbf{99.9} \%  \\
\bottomrule\\
\end{tabular}%
\caption{Results (including posterior median, 95\% High Density Intervals, and Probability of Direction) from study 1 and 2 on Perceived Surprise.}
\label{tab:studiesSurprise}
\end{table}

% \begin{figure}[tb]% specify a combination of t, b, p, or h for top, bottom, on its own page, or here
%   \centering % avoid the use of \begin{center}...\end{center} and use \centering instead (more compact)
%   \includegraphics[width=\columnwidth]{figs/study2/Surprise-02.pdf}
%   \caption{Cumulative frequency of responses to the interest, recommend, and surprise questions for Study 2.
%   }
%   \label{fig:study2-engagement}
% \end{figure}

\textbf{Study 2}: %This indicated that perceived surprise was higher in the Draw condition compared to the Control condition (Draw -- Control = .62, 95\% HDI [.26, .99], $PD = 99.9\%$) and the Categorize condition (Draw -- Categorize = .66, 95\% HDI [.30, 1.05], $PD = 100\%$), while there was no difference between the Categorize and Control conditions (Categorize -- Control = -.04, 95\% HDI [-.40, .33], $PD = 57.5\%$).
%\textcolor{revision2}
\revAliFinal{The results indicated participants in the Draw condition were on average more likely to indicate higher perceived surprise when compared to the Control and the Categorize conditions, while there was no difference between the Categorize and Control conditions, Table \ref{tab:studiesSurprise}b.} Consistent with the findings from Study 1, there were no effects for the perceived interest and recommend questions. Looking at the \revDougFinal{distribution of} responses to the surprise question (Figure \ref{fig:study1-engagement}, bottom right), more participants in the Draw condition found the articles ``extremely" surprising while fewer participants responded ``Not at all" surprising. 
%Compared to the individual responses on the surprise question from study 1, we observe a shift in participants to choosing a higher point in the 5-point Likert scale.

Overall, the results from both studies showed a consistent increase in \textit{perceived surprise} for the Draw condition compared to both the Control and Categorize conditions, while there was no evidence for differences for the other engagement questions.

\section{\textcolor{black}{H2}: The impact of \textcolor{revision2}{contrasting narratives} on recall and engagement}

\begin{figure}[ht]% specify a combination of t, b, p, or h for top, bottom, on its own page, or here
  \centering % avoid the use of \begin{center}...\end{center} and use \centering instead (more compact)
  \includegraphics[width=\columnwidth]{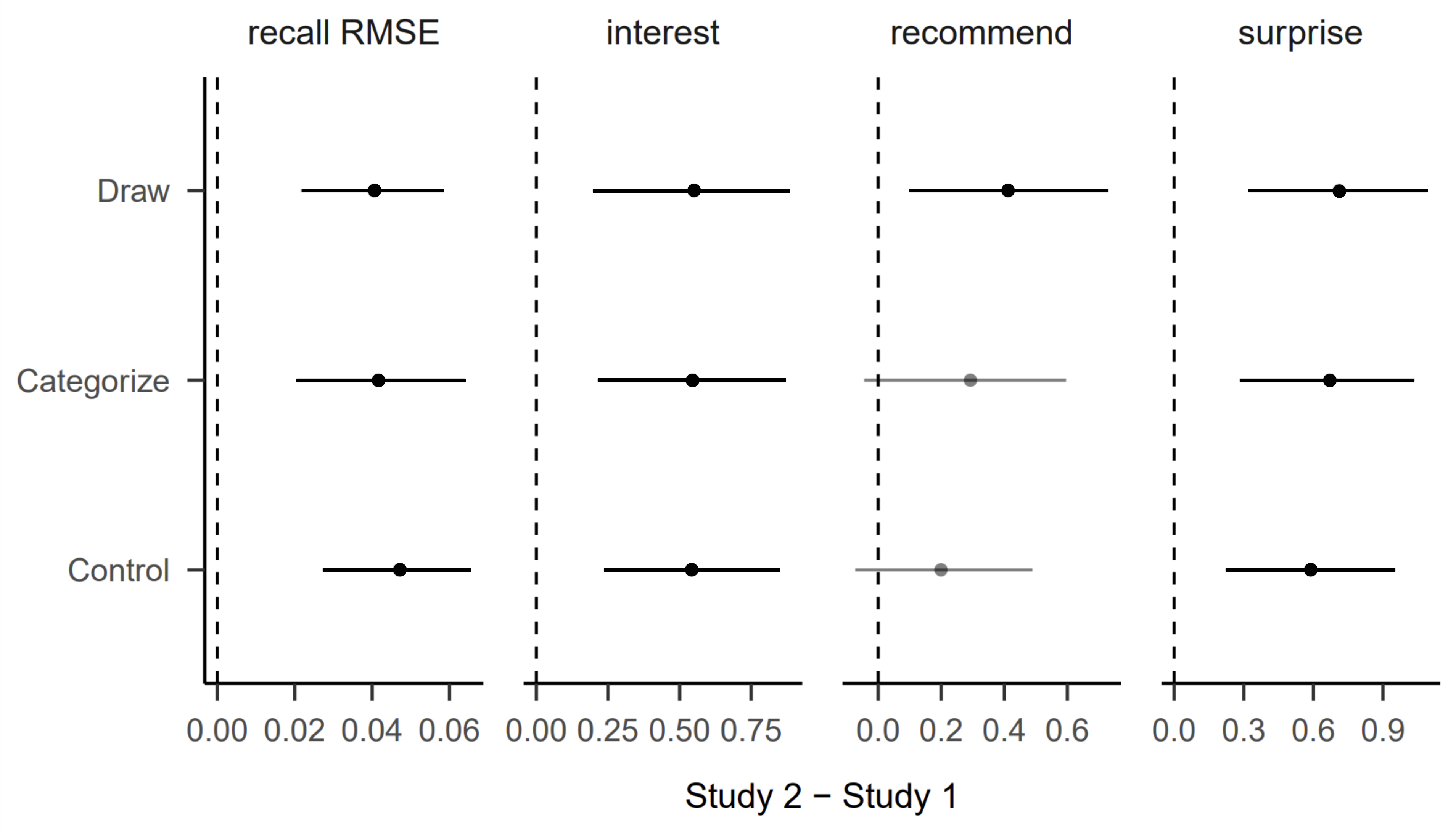}
  \caption{Contrasts between Study 2 and Study 1 for models of recall RMSE and engagement.
  }
  \label{fig:h3contrasts}
\end{figure}

\begin{figure*}[ht]% specify a combination of t, b, p, or h for top, bottom, on its own page, or here
  \centering 
  \includegraphics[width=6in]{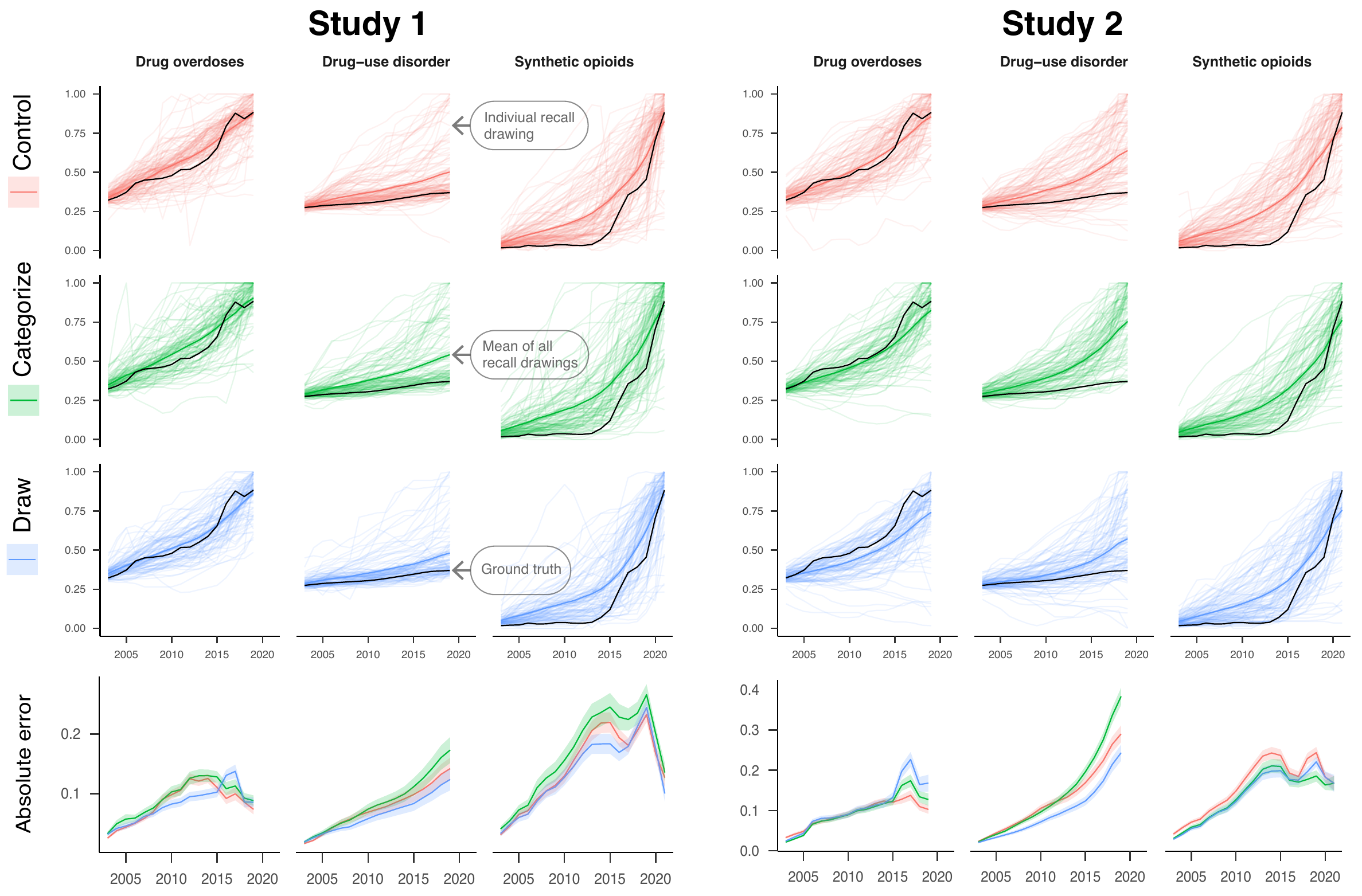}
  \caption{\textbf{Top:} Responses and actual trend (black line) in Study 1 \& 2. The trend values are scaled according to how they looked to the participants during the studies. \textbf{Bottom:} Yearly absolute error (mean and standard error) in Study 1 \& 2.
  }
  \label{fig:recall}
\end{figure*} 

In this set of analyses, we examined the effect of \textcolor{revision2}{contrasting narratives} on recall and engagement for the articles by directly comparing the effects from Study 1 and Study 2.
Given the similarity of the designs of Study 1 and Study 2, including the same set of outcome measures, we performed an integrative analysis which combined the datasets of the two studies \cite{curran_integrative_2009}. \revDougFinal{Integrative data analysis refers to strategies in which two or more independent data sets are pooled or combined into one and then statistically analyzed. It offers advantages over meta-analysis in which \textit{summary statistics} across multiple studies are pooled together \cite{curran_integrative_2009}.}

We pooled the data for both studies and fit new Bayesian models for each outcome which included the study (1 vs. 2) \revDougFinal{and the treatment $\times$ study interaction} as predictors.
\revDougFinal{The model specifications were otherwise identical to those used for the analyses of the individual studies above, with a baseline model (\texttt{y $\sim$ study * treatment + article + (1|PID)}) and alternative model (\texttt{y $\sim$ study * treatment + article + 
topicInvolvement + preAttitude + (1|PID)}).}
After fitting the models we calculated, within each treatment, the contrast between studies to assess the impact of \textcolor{revision2}{contrasting narratives} (Study 2) relative to when they are not presented (Study 1) (see Figure \ref{fig:h3contrasts}).
\revDoug{It should be noted that there may be other confounding differences between the two studies which naturally limit our conclusions from this analysis, including potential differences in samples that were collected at different timepoints. We discuss these potential limitations further in Section \ref{sec:discussion}}.

% As in Section \ref{sec:elicitation_impact}, we used the combined datasets for both studies and fit new models for each outcome with study (1 vs. 2), treatment (Draw, Categorize, Control), and their interaction as predictors.
% We then calculated, within each treatment, the contrast between studies to assess the impact of \textcolor{revision2}{contrasting narratives} (Study 2) relative to when they are not presented (Study 1) (see Figure \ref{fig:h3contrasts}).

\begin{figure}[ht]% specify a combination of t, b, p, or h for top, bottom, on its own page, or here
  \centering % avoid the use of \begin{center}...\end{center} and use \centering instead (more compact)
  \includegraphics[width=\columnwidth]{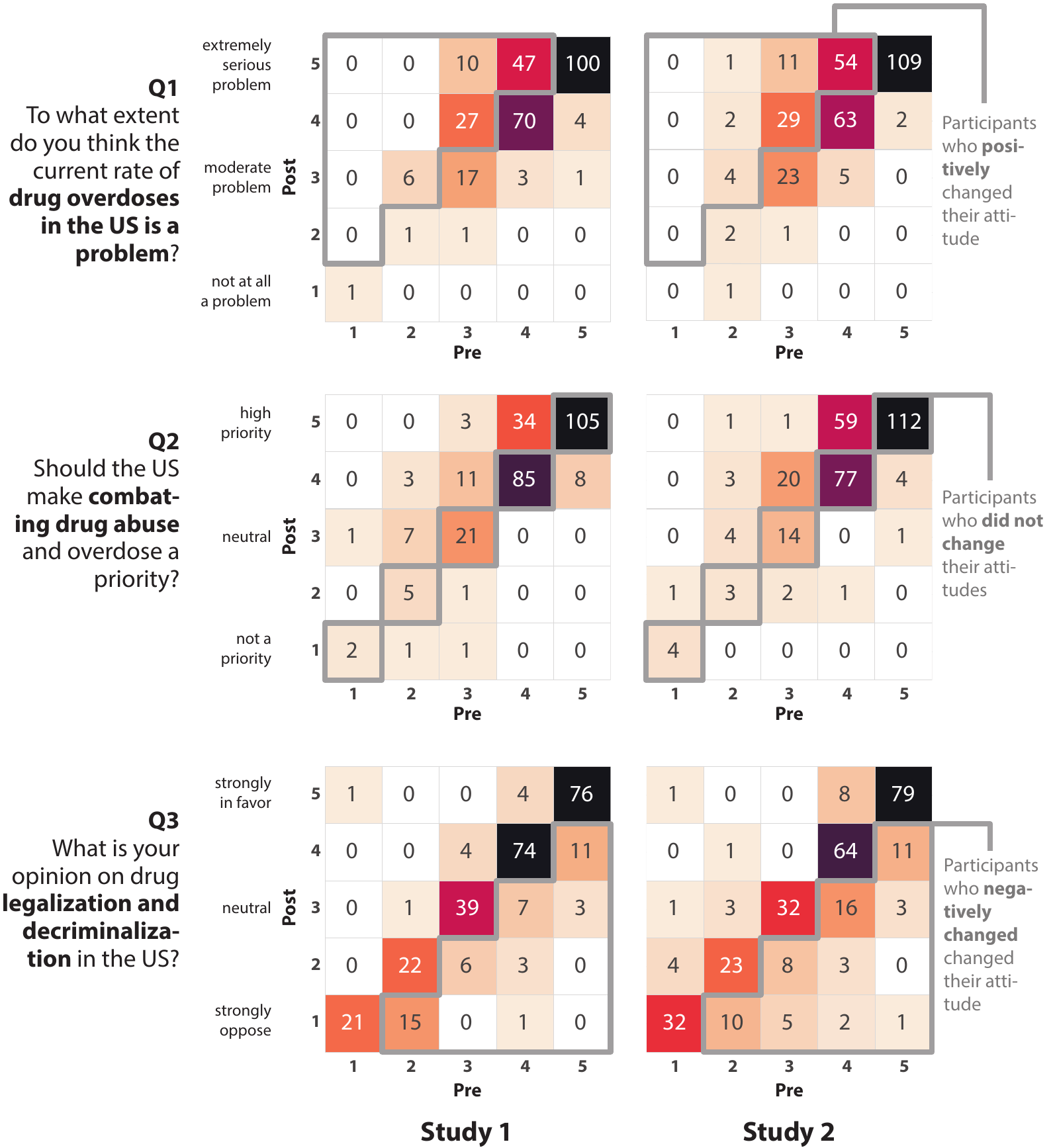}
  \caption{\textcolor{black}{ Heatmap visualization of participants' attitude responses before and after completing the study \textbf{Left:} Pre and post response to attitude questions in study 1. \textbf{Right:} Pre and post response to attitude questions in study 2. }
  }
  \label{fig:attitudeRawchange}
\end{figure}

\subsection{Recall error}

\revAliFinal{The results for recall accuracy indicated that participants on average were likely to be less accurate in recreating the trends in Study 2 compared to Study 1, }\textcolor{revision}{indicating the contrasting narratives had a negative impact on recall accuracy} \revAliFinal{ potentially due to being exposed to multiple datasets in a single article.} 
%As shown in the first column in Fig \ref{fig:h3contrasts}, recall RMSE was higher in Study 2 across all treatment conditions: Control (Study 2 -- Study 1 = .04, 95\% HDI [.03, .06], $PD = 100\%$), Categorize (Study 2 -- Study 1 = .04, 95\% HDI [.02, .06], $PD = 100\%$), and Draw (Study 2 -- Study 1 = .04, 95\% HDI [.02, .06], $PD = 100\%$) conditions.
\revDougFinal{As shown in Figure \ref{fig:recall-rmse}},
% As shown in the first column in Fig \ref{fig:h3contrasts}, 
recall RMSE was higher in Study 2 across all treatment conditions (Table \ref{tab:contrastingNarrativesRecallRMSE}).
Figure \ref{fig:recall} right shows bigger gaps/discrepancies between the mean of recall drawings in each treatment and article compared to corresponding results from Study 1. The aggregate absolute error on the bottom also showed higher error from Study 2 (the error range on the y axis is much larger in Study 2). 
% We discuss higher recall error found in study 2 in section \ref{sec:discussion}.

\begin{table}[htb]
\centering
\footnotesize
\begin{tabular}{@{}ccll@{}}
\toprule
\multicolumn{4}{c}{\textbf{Differences in Recall RMSE Between Study 1 and 2}}                            \\ \midrule
\rowcolor[HTML]{EFEFEF} 
\multicolumn{1}{l}{\cellcolor[HTML]{EFEFEF}\textbf{Treatment}} &
  \textbf{Study2 - Study1} &
  \multicolumn{1}{c}{\cellcolor[HTML]{EFEFEF}\textbf{95\% HDI}} &
  \multicolumn{1}{c}{\cellcolor[HTML]{EFEFEF}\textbf{PD}} \\ \midrule
% \multicolumn{4}{c}{\textbf{a: Study 1}}                            \\ \midrule

\textbf{Control}    & \textbf{0.04 }                    & \textbf{{[}0.02 , 0.06{]}} & \textbf{100\%}   \\
\textbf{Categorize} & \textbf{0.04}                      & \textbf{{[}0.02 , 0.06{]}}    & \textbf{100 \%}   \\
\textbf{Draw}       & \textbf{0.04}                     & \textbf{{[}0.02 , 0.06{]}}   & \textbf{100\% }  \\ \midrule

\end{tabular}%

\caption{Differences in recall RMSE between Study 1 and 2.}
\label{tab:contrastingNarrativesRecallRMSE}
\end{table}

\begin{table}[htb]
\centering
\footnotesize
\begin{tabular}{@{}ccll@{}}
\toprule
\multicolumn{4}{c}{\textbf{Impact of Contrasting Narratives on Article Engagement}}                            \\ \midrule
\rowcolor[HTML]{EFEFEF} 
\multicolumn{1}{l}{\cellcolor[HTML]{EFEFEF}\textbf{Treatment}} &
  \textbf{Study2 - Study1} &
  \multicolumn{1}{c}{\cellcolor[HTML]{EFEFEF}\textbf{95\% HDI}} &
  \multicolumn{1}{c}{\cellcolor[HTML]{EFEFEF}\textbf{PD}} \\ \midrule
\multicolumn{4}{c}{\textbf{a: Surprise}}                            \\ \midrule

\textbf{Control}    & \textbf{0.52}                     & \textbf{{[}0.17 , 0.89{]}}  & \textbf{99.9\%}   \\
\textbf{Categorize} & \textbf{0.63}                     & \textbf{{[}0.25 , 1.00{]}} & \textbf{99.9\%}   \\
\textbf{Draw}       & \textbf{0.80}                     & \textbf{{[}0.43 , 1.17{]}}  & \textbf{100\%}   \\ \midrule

\multicolumn{4}{c}{\textbf{b: Interest}}                            \\ \midrule

\textbf{Control}    & \textbf{0.49}                     & \textbf{{[}0.22 , 0.76{]}}  & \textbf{100\%}   \\
\textbf{Categorize} & \textbf{0.46 }                    & \textbf{{[}0.16 , 0.75{]}}  & \textbf{99.9\%}   \\
\textbf{Draw}       & \textbf{0.67}                     & \textbf{{[}0.39 , 0.97{]}}  & \textbf{100\%}   \\

\midrule
\multicolumn{4}{c}{\textbf{c: Recommend}}                            \\ \midrule

Control    & 0.17                     & {[}-.07 , 0.40{]}  & 92.3\%   \\
Categorize & 0.22                     & {[}-.04 , 0.49{]}  & 94.5\%   \\
\textbf{Draw}       & \textbf{0.52}                     & \textbf{{[}0.25 , 0.79{]}}  & \textbf{100\%}   \\
\bottomrule\\
\end{tabular}%

\caption{The impact of contrasting narratives on article engagement. }
\label{tab:contrastingNarrativesArticleEngagement}
\end{table}

\subsection{Article engagement}

We found substantial evidence of increased interest and surprise across all treatment conditions in Study 2 relative to Study 1 (Table \ref{tab:contrastingNarrativesArticleEngagement}a,b), suggesting that the addition of \textcolor{revision2}{contrasting narratives} \revAliFinal{ was more likely to create a positive impact on participant engagement with the articles. }
Responses to the Recommend question were higher in Study 2 for the Draw condition, but there was less evidence of a credible difference between studies in the Control and Categorize conditions (Table \ref{tab:contrastingNarrativesArticleEngagement}c).

% Surprise was higher in Study 2 for the Control (Study 2 -- Study 1 = .59, 95\% HDI [.22, .95], $PD = 99.9\%$), Categorize (Study 2 -- Study 1 = .67, 95\% HDI [.28, 1.04], $PD = 99.9\%$), and Draw (Study 2 -- Study 1 = .71, 95\% HDI [.32, 1.09], $PD = 100\%$) conditions.

% Responses to the Interest question were higher in Study 2 for the Control (Study 2 -- Study 1 = .54, 95\% HDI [.24, .85], $PD = 100\%$), Categorize (Study 2 -- Study 1 = .54, 95\% HDI [.21, .87], $PD = 99.9\%$), and Draw (Study 2 -- Study 1 = .55, 95\% HDI [.20, .88], $PD = 99.9\%$) conditions.

% Responses to the Recommend question were higher in Study 2 for the Draw condition (Study 2 -- Study 1 = .41, 95\% HDI [.10, .60], $PD = 99.6\%$).
% There was weaker evidence for increases in the 
% Categorize condition (Study 2 -- Study 1 = .29, 95\% HDI [-.04, 1.60], $PD = 96.3\%$) and the Control condition (Study 2 -- Study 1 = .20, 95\% HDI [-.07, .49], $PD = 92.5\%$) conditions.

% \begin{figure}[h]% specify a combination of t, b, p, or h for top, bottom, on its own page, or here
%   \centering % avoid the use of \begin{center}...\end{center} and use \centering instead (more compact)
%   \includegraphics[width=\columnwidth]{figs/heatmap_test.pdf}
%   \caption{\textcolor{black}{ Heatmap visualization of participants' attitude responses before and after completing the study \textbf{Left:} Pre and post response to attitude questions in study 1. \textbf{Right:} Pre and post response to attitude questions in study 2. }
%   }
%   \label{fig:attitudeRawchange}
% \end{figure}

\section{\textcolor{black}{H3: Results on Attitude Change}}
\subsection{Overall attitude change}

\begin{figure}[ht]% specify a combination of t, b, p, or h for top, bottom, on its own page, or here
  \centering % avoid the use of \begin{center}...\end{center} and use \centering instead (more compact)
  \includegraphics[width=\columnwidth]{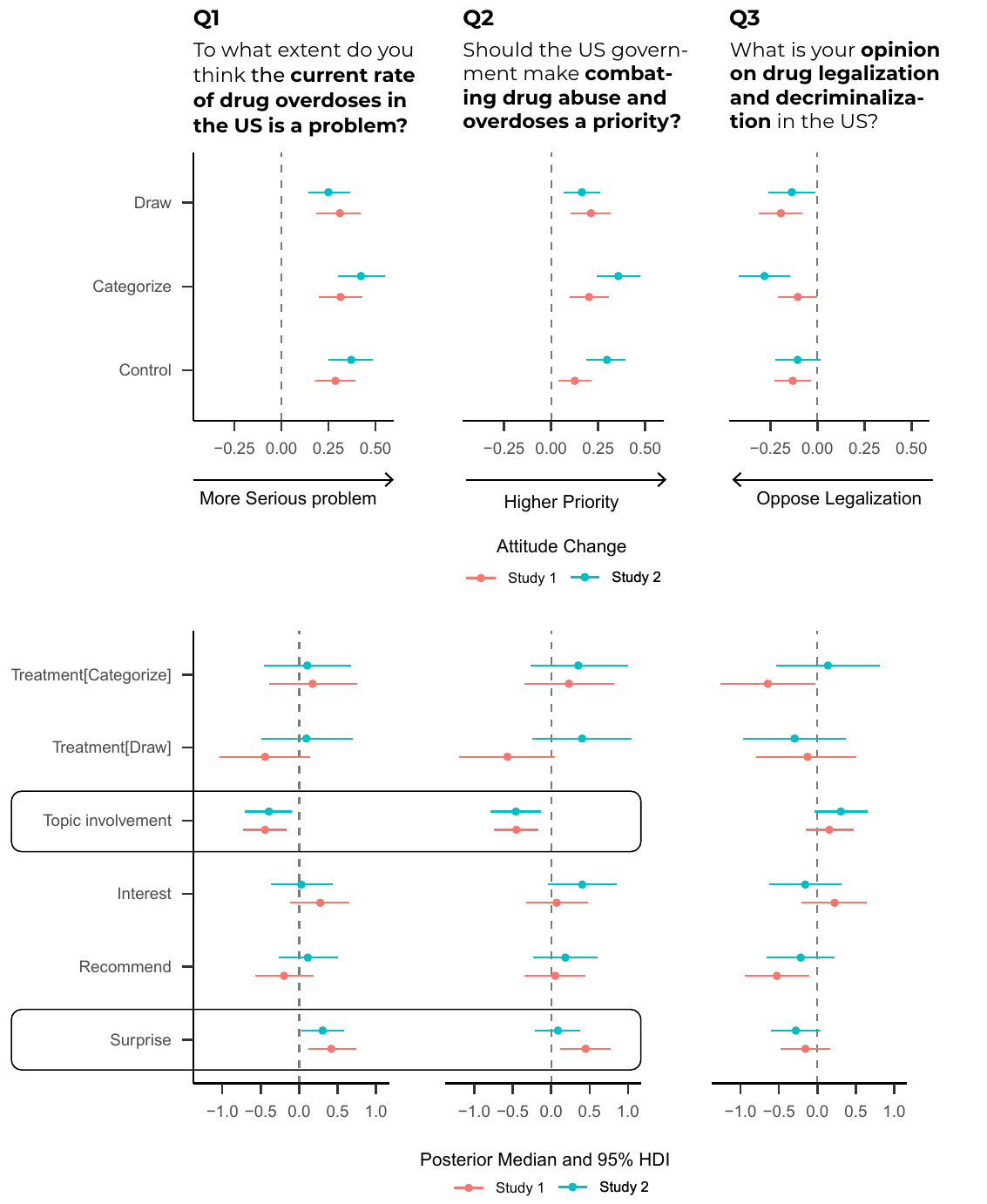}
  \caption{\textbf{Top:} Attitude change measured by 3 questions for each treatment condition (predicted means and 95\% HDIs). \textbf{Bottom:} Posterior medians and 95\% HDIs for attitude change model coefficients.
  }
  \label{fig:attitudechange}
\end{figure}

To measure attitude change, we calculated the difference between responses to the three attitude questions before and after completing the tasks for each study. 
Since attitude change scores were calculated as the difference between ratings on 5-point scales the resulting distribution was discrete and highly skewed.
We therefore fit ordinal regression models (with cumulative logit link functions) to attitude change scores for each question.
For both studies, we first fit a baseline model with elicitation treatment as the only additional predictor besides the thresholds between response categories.
We also explored a more complex model in which we added topic involvement and average responses to the article engagement questions (interest, recommend, and surprise) as predictors.
\revDoug{Model comparisons with LOO-PSIS favored the more complex model.}
Figure \ref{fig:attitudechange} (bottom) shows the posterior median and 95\% HDIs of the focal parameters for the attitude change model in both studies. 

For Q1 (\textit{To what extent do you think the current rate of drug overdoses in
the US is a problem?}) and Q2 (\textit{Should the US government make combating drug abuse and
overdoses a priority?}), positive changes in responses suggest increased perceived importance of the drug overdose issue.
For Q3 (\textit{What is your opinion on drug legalization and decriminalization
in the US}), positive changes indicate increased support for legalization/decriminalization while negative changes indicate decreased support. \footnote{\textcolor{revision}{We pre-registered to derive a compound score for measuring attitude change. Upon examination, we agreed that Q3 captures more complex policy-related aspects of attitude towards drug overdose compared to the first two questions. Therefore, we report the analysis of the three attitude questions separately.}}
\textcolor{revision}{Overall, we see mostly positive attitude change for Q1, with almost 1/3 of participants in both Study 1 and 2 shifting their attitude to perceiving drug overdose as a more serious problem.
In comparison, a smaller proportion of participants shifted their opinion on Q2 and Q3. Figure \ref{fig:attitudechange} shows model predictions for attitude change and Figure \ref{fig:attitudeRawchange} shows the pre and post-responses for each attitude question from both studies.}

\textcolor{revision}{To report the results on the impact of elicitation and contrasting narratives on attitude change,} we first elaborate on the questions used to measure attitude change and show summary statistics on the overall attitude change.

%However, the directions of the shift were systematic on all three questions with the majority of participants reporting positive shifts for questions 1\&2 and negative shifts for question 3. 
%As is made apparent by the concentration of gray circles in the top right corner of all charts in Figure 5, many participants were already reporting the highest scale in the pre-questionnaire, leaving little room for attitude change. 

\revAli{While more participants overall did not change their attitudes, the directions of shift for the ones who did were systematic on all three questions. For questions 1\&2, aside from participants with the most extreme pre-attitude choices with no room for positive attitude change (i.e., extremely serious problem for Q1 and high priority for Q2), the majority of participants had a positive attitude change (See Figure \ref{fig:attitudeRawchange} top and middle rows). For question 3, where overall participants' pre-attitudes were more spread out, we observe many participants shifting attitudes towards opposition to drug legalization (See Figure \ref{fig:attitudeRawchange} bottom row). While we do believe a general shift in attitudes for participants who had room to change in the provided scale is interesting, it is important to note that although we do observe general shifts in attitudes for many participants, these changes at different parts of the scale might not be equal. For example, a shift from somewhat opposed to neutral might have a completely different weight in comparison to one from neutral to somewhat in favor. Future studies should dive deeper into the nuances of attitude change.}

\revMilad{We contend the overall observation on attitude change is related to our selection of a real-world topic that people have thought of and may have already formed an opinion. In comparison to the uncontentious topics chosen in other studies \cite{karduni2020bayesian}, we instead picked a topic that resembles many issues we wrangle with in our daily lives where most of us have formed our opinions and see if there is room for data visualizations to persuade.} \revAliFinal{Moreover, We intentionally refrained from eliciting participants' attitudes after each article to minimize fatigue and because all articles covered the same topic (Drug abuse) and therefore this study cannot shed light on whether introducing only one article might produce similar effects.}

\begin{table}[htb]
\centering
\footnotesize
\begin{tabular}{@{\hspace{1pt}}c@{\hspace{1pt}}@{\hspace{1pt}}c@{\hspace{1pt}}@{\hspace{1pt}}c@{\hspace{1pt}}@{\hspace{1pt}}l@{\hspace{1pt}}@{\hspace{1pt}}c@{\hspace{1pt}}}
\toprule
\multicolumn{5}{c}{\textbf{Treatment Pairwise Attitude Differences}} \\ 
\midrule
\multicolumn{5}{c}{\textbf{a: Study 1}} \\ 
\midrule
\rowcolor[HTML]{EFEFEF} 
\multicolumn{1}{l}{\cellcolor[HTML]{EFEFEF}\textbf{Questions}} &
  \textbf{Treatment Pairs} &
  \textbf{Attitude Change} &
  \multicolumn{1}{c}{\cellcolor[HTML]{EFEFEF}\textbf{95\% HDI}} &
  \multicolumn{1}{c}{\cellcolor[HTML]{EFEFEF}\textbf{PD}} \\
                              & Draw - Control       & 0.05   & {[}-0.13, 0.22{]} & 70\% \\
                              & Draw - Categorize    & 0.02 & {[}-0.16, 0.20{]}  & 57.8\% \\
\multirow{-3}{*}{Q1} & Categorize - Control & 0.03   & {[}-0.15, 0.20{]}  & 63.1\% \\
\multicolumn{5}{l}{}                                                                                \\
                              & Draw - Control      & 0.07   & {[}-0.07, 0.24{]}  & 83.3\% \\
                              & Draw - Categorize    & -0.01   & {[}-0.18, 0.17{]}  & 53\% \\
\multirow{-3}{*}{Q2} & Categorize - Control & 0.08   & {[}-0.07, 0.24{]}  & 85.6\%   \\
\multicolumn{5}{l}{}                                                                                \\
                              & Draw - Control       & -0.06  & {[}-0.22, 0.11{]} & 66.1\% \\
                              & Draw - Categorize    & -0.09  & {[}-0.25, 0.08{]}  & 85.3\% \\
\multirow{-3}{*}{Q3} & Categorize - Control & 0.03   & {[}-0.12, 0.18{]}  & 66.1\% \\ 
\midrule
\multicolumn{5}{c}{\textbf{b: Study 2}} \\ 
\midrule
\rowcolor[HTML]{EFEFEF} 
\multicolumn{1}{l}{\cellcolor[HTML]{EFEFEF}\textbf{Questions}} &
  \textbf{Treatment Pairs} &
  \textbf{Attitude Change} &
  \multicolumn{1}{c}{\cellcolor[HTML]{EFEFEF}\textbf{95\% HDI}} &
  \multicolumn{1}{c}{\cellcolor[HTML]{EFEFEF}\textbf{PD}} \\
                              & Draw - Control       & -0.12 & {[}-0.28, 0.04{]}   & 92.8\%  \\
                              & Draw - Categorize    & -0.17 & {[}-0.33, 0.01{]}   & 98.1\%  \\
\multirow{-3}{*}{Q1} & Categorize - Control & 0.05  & {[}-0.11, 0.23{]}   & 73.2\%    \\
\multicolumn{5}{l}{}                                                                                  \\
                              & Draw - Control      & -0.13 & {[}-0.28, 0.01{]}   & 96.8\%  \\
                              & \textbf{Draw - Categorize}    & \textbf{-0.19} & \textbf{{[}-0.35, -0.05{]}}  & \textbf{99.3\%}  \\
\multirow{-3}{*}{Q2} & Categorize - Control & 0.06  & {[}-0.09, 0.22{]}   & 77.6\%  \\
\multicolumn{5}{l}{}                                                                                  \\
                              & Draw - Control       & -0.03 & {[}-.21, 0.13{]}    & 64.3\% \\
                              & Draw - Categorize    & 0.15  & {[}-0.04, 0.33{]}   & 94.2\%  \\
\multirow{-3}{*}{Q3} & \textbf{Categorize - Control} & \textbf{-0.17} & \textbf{{[}-0.35, -0.01{]}} & \textbf{97.6\%}  \\ 

\cmidrule(l){1-5} 
\end{tabular}%

\caption{Treatment pairwise attitude difference in Studies 1 and 2. }
\label{tab:study1AttitudeChange}
\end{table}

\subsection{Impact of belief elicitation on attitude change}

% \begin{figure}[t]% specify a combination of t, b, p, or h for top, bottom, on its own page, or here
%   \centering % avoid the use of \begin{center}...\end{center} and use \centering instead (more compact)
%   % \includegraphics[width=\columnwidth]{figs/attitude_change_combined.pdf}
%   \includegraphics[width=\columnwidth]{figs/attitude_change_combined_new.pdf}
%   \caption{\textbf{Top:} Attitude change measured by 3 questions for each treatment condition (predicted means and 95\% HDIs). \textbf{Bottom:} Posterior medians and 95\% HDIs for attitude change model coefficients.
%   }
%   \label{fig:attitudechange}
% \end{figure}

Figure \ref{fig:attitudechange} (top) shows the expected posterior median and 95\% HDI for the amount of attitude change for each question.
Overall changes in responses to questions 1 and 2 were broadly positive in both Study 1 and Study 2, while changes were in the negative direction for question 3, leaning more against drug legalization.
This indicates that there were small but reliable changes in attitude about the topic as a result of viewing the charts.

\revDoug{We first used the estimated model} to examine whether the amount of attitude change differed between treatment conditions.
For each attitude question (Q1, Q2, and Q3) we calculated the pairwise contrasts between treatment conditions 
% (see Figure \ref{fig:attitudechange}, top) 
for each study (Table \ref{tab:study1AttitudeChange}).
\textcolor{black}{For \textbf{Study 1-Elicitation}, we did not observe an impact of elicitation on the amount of attitude change for Q1, Q2, or Q3.}
% , see Table \ref{tab:study1AttitudeChange}a.}
% Thus attitudes appeared to shift to a similar extent across the conditions with no additional effects of elicitation type.
\textcolor{black}{The results for \textbf{Study 2-Elicitation+Contrasts} were generally consistent with Study 1 in showing similar amounts of attitude change across treatments (Table \ref{tab:study1AttitudeChange}b).} 
There were no differences in attitude change for Q1. 
For Q2, attitude changes were smaller in the Draw condition compared to the Categorize condition, while for Q3 attitude change was more negative in the Categorize condition compared to the Control condition.
In sum, while there were some indications of small differences in attitude change between treatments, no consistent effects of elicitation method were observed.

% \begin{figure}[t]% specify a combination of t, b, p, or h for top, bottom, on its own page, or here
%   \centering % avoid the use of \begin{center}...\end{center} and use \centering instead (more compact)
%   % \includegraphics[width=\columnwidth]{figs/attitude_change_combined.pdf}
%   \includegraphics[width=\columnwidth]{figs/attitude_change_combined_new.pdf}
%   \caption{\textbf{Top:} Attitude change measured by 3 questions for each treatment condition (predicted means and 95\% HDIs). \textbf{Bottom:} Posterior medians and 95\% HDIs for attitude change model coefficients.
%   }
%   \label{fig:attitudechange}
% \end{figure}

\subsection{Impact of engagement on attitude change}

The estimated model also indicated effects of topic involvement and engagement with the articles on attitude change (Figure \ref{fig:attitudechange}, bottom).
First, greater topic involvement was associated with smaller changes in attitude for questions 1 and 2. 
This is likely because topic involvement was positively correlated with initial responses to these questions, implying that participants who expressed high involvement already gave the maximal rating to these questions.
Second, we found that average perceived surprise to the articles was associated with positive changes in responses to Q1 (Studies 1 and 2) and Q2 (Study 1), while the likelihood of recommending the article was associated with more negative changes in responses to Q3 (Study 2). 
\revDoug{Together, these results suggest that participants who were less involved in the topic and who were more surprised by the trends were more likely to subsequently express greater concern compared to the beginning of the study.}

\subsection{Impact of \textcolor{revision2}{contrasting narratives} on attitude change}

% In order to assess the effect of \textcolor{revision2}{contrasting narratives} on attitude change we combined the datasets from Studies 1 and 2 and fit a new ordinal regression model for each attitude question.
% The model included study (1 vs. 2), treatment (Draw, Categorize, Control) and their interaction as predictors.
% For each question (Q1, Q2, and Q3) we then calculated the expected difference in attitude change between the two studies within each treatment condition.

% For Q1 there was no evidence for differences between Study 1 and Study 2 (Draw = -.04, 95\% HDI [-.19, .13], $PD = 69.2\%$;
% Categorize = .06, 95\% HDI [-.11, .24], $PD = 74.9\%$;
% Control = .05, 95\% HDI [-.10, .22], $PD = 77.1\%$).

% For Q2 there was evidence for greater attitude change in the Control condition for Study 2 compared to Study 1 (Control = .17, 95\% HDI [.03, .31], $PD = 98.8\%$) but not the other conditions (Draw = .01, 95\% HDI [-.14, .15], $PD = 53.0\%$;
% Categorize = .13, 95\% HDI [-.02, .29], $PD = 94.7\%$).

% For Q3 there was no evidence for differences between Study 1 and Study 2 (Draw = .07, 95\% HDI [-.11, .24], $PD = 77.1\%$;
% Categorize = -.16, 95\% HDI [-.34, .22], $PD = 96.9\%$;
% Control = .06, 95\% HDI [-.09, .22], $PD = 78.6\%$).

% Thus, these comparisons suggest that while participants in study 2 exhibited overall attitude updates, the addition of \textcolor{revision2}{contrasting narratives} in Study 2 did not lead to a significant amount of attitude change relative to Study 1.

\textcolor{black}{In order to assess the effect of \textcolor{revision2}{contrasting narratives} on attitude change we combined the datasets from Studies 1 and 2 and fit a new Bayesian ordinal regression model for each attitude question.
The model included study (1 vs. 2), treatment (Draw, Categorize, Control) and their interaction as fixed effects.
For each question (Q1, Q2, and Q3) we then calculated the expected difference in attitude change between the two studies within each treatment condition.
For Q1 and Q3 there was no evidence for differences between Study 1 and Study 2. For Q2 there was evidence for greater attitude change in the Control condition for Study 2 compared to Study 1, but not the other conditions, see Table \ref{tab:AttitudeChangeStudies}.
Thus, these comparisons suggest that while participants in study 2 exhibited overall attitude changes, the addition of \textcolor{revision2}{contrasting narratives} in Study 2 did not lead to a significant amount of attitude change relative to Study 1.}

\begin{table}[t]
\centering
\footnotesize
\begin{tabular}{@{}cccll@{}}
\toprule
\multicolumn{5}{c}{\textbf{Impact of \textcolor{revision2}{contrasting narratives} on Attitude Change}}                 \\ \midrule
\rowcolor[HTML]{EFEFEF} 
\multicolumn{1}{l}{\cellcolor[HTML]{EFEFEF}\textbf{Questions}} &
  \multicolumn{1}{l}{\cellcolor[HTML]{EFEFEF}\textbf{Treatment}} &
  \textbf{Attitude Change} &
  \multicolumn{1}{c}{\cellcolor[HTML]{EFEFEF}\textbf{95\% HDI}} &
  \multicolumn{1}{c}{\cellcolor[HTML]{EFEFEF}\textbf{PD}} \\
                              & Draw       & -0.06 & {[}-0.23, 0.11{]}  & 75.6\% \\
                              & Categorize & 0.08  & {[}-0.10, 0.25{]}  & 80.7\% \\
\multirow{-3}{*}{Q1} & Control    & 0.05  & {[}-0.11, 0.22{]}  & 75.5\% \\
\multicolumn{5}{l}{}                                                                      \\
                              & Draw       & 0.01  & {[}-0.13, 0.17{]}  & 59.8\%   \\
                              & Categorize & 0.13  & {[}-0.03, 0.29{]} & 94.1\% \\
\multirow{-3}{*}{\textbf{Q2}} & \textbf{Control}    & \textbf{0.16}  & \textbf{{[}0.01, 0.30{]}}   & \textbf{98.5\%} \\
\multicolumn{5}{l}{}                                                                      \\
                              & Draw       & 0.05  & {[}-.14, 0.22{]}   & 71\% \\
                              & Categorize & -0.16 & {[}-0.34, 0.00{]}  & 97.1\% \\
\multirow{-3}{*}{Q3} & Control    & 0.05  & {[}-0.14, 0.22{]}  & 76\% \\ \cmidrule(l){2-5} 
\end{tabular}%

\caption{Impact of \textcolor{revision2}{contrasting narratives} on attitude change between studies. }
\label{tab:AttitudeChangeStudies}
\end{table}

%\subsubsection{Attitude change}

% \subsection{Discussion of \textcolor{revision2}{contrasting narratives} impact}

\section{Participants}
\label{sec:participants}

We recruited all participants from Prolific, limited to participants in the United States and above 95\% job approval rate.
\subsection{Study 1: Elicitation}
We recruited 306 participants from Prolific. Participants earned \$4.5 for completing the task, which on average took 13.8 minutes (SD = 7.8). Per our preregistration, we excluded 1 participant who did not complete the full study. We planned to exclude any participants who gave non-sensible or inappropriate responses to the open-ended questions for the CRT, but no participants met this criterion. We followed our pre-registration and excluded participants who complete the study below or above two standard deviations of average completion time. %However, we deviated from this exclusion criterion based on completion time because of the variation. The completion time for Study one ranges from 3.1 to 55.2 minutes with a median of 11.7 minutes.

After excluding responses based on incompleteness and time, there were $N$ = 288 participants included in the analyses (Control: 106; Categorize: 91; Draw: 91). The average age was 35.6 years, range: 18–78). 110 identified as female, 166 identified as male, 8 as other, and 4 chose 'prefer not to say'. 194 were White, 26 were Black/African-American, 39 were Asian, 17 were Hispanic, 1 was Middle Eastern, 8 were another race/ethnicity, and 3 chose 'prefer not to say'. 143 participants reported a college degree, 87 with a high school degree, 36 with a Masters's degree, 7 with a Doctorate degree and 15 chose either ``other" or ``prefer not to say".

\subsection{Study 2: Elicitation+Contrasts}
We recruited 320 participants from Prolific. One response was excluded due to incompleteness. Participants earned \$6 for completing the task, which on average took 19.1 minutes (SD = 11.9). Per our preregistration, we planned to exclude any participants who gave non-sensible or inappropriate responses to the open-ended questions for the CRT, but no participants met this criterion. We also pre-registered to exclude participants whose completion time is above or below 3 standard deviations. This criterion resulted in 12 exclusions. The completion times ranges from 5.3 to 82 minutes, with a median of 15.5 minutes.

Among $N$ = 307 participants for study 2,  110 were in the Control condition, 98 were in the categorize trend condition, while 99 were in the draw trend condition. The average age was 38.4 years, range: 18–79). 148 identified as female, 151 identified as male, 3 as other, and 5 chose 'prefer not to say'. 224 were White, 19 were Black/African-American, 20 were Asian, 22 were Hispanic, 4 were American Indian or Alaska Native, 11 were another race/ethnicity, and 7 chose 'prefer not to say.' 130 participants had obtained a college degree, 109 reported a High School degree, 36 with a Masters's degree, 7 with a Doctorate degree, and 25 chose either ``other" or ``prefer not to say".

\section{Discussion, Limitations, and Future Work}
\label{sec:discussion}

\subsection{The persuasive power of journalistic articles with data visualization}

Journalistic visualizations often aim to influence attitudes and behaviors. 
We examined how belief elicitation and charts with contrasting trends affect attitude change. 
We created interactive articles similar to the NYTimes article ``You Draw It: Just How Bad Is the Drug Overdose Epidemic?" \cite{YouDrawIt}. 
We observed systematic changes in participants' attitudes \revDoug{across three dimensions: After viewing the visualizations, participants expressed more concern about drug overdoses, more support for government intervention, and more opposition to drug legalization/decriminalization. While these changes in attitude were relatively small, they are nevertheless notable since past work suggests that} achieving attitude change is often difficult \cite{Markant_2023, Heyer_20, Pandey_2014, Liem_2020} \revDoug{and because many participants' preexisting attitudes were already strong, leaving relatively little room for change}.
In most cases, participants already believed that the drug epidemic was a problem and should be a priority; after interacting with the visualizations, they believed it was even more important (Figure \ref{fig:attitudeRawchange}).

\revMiladFinal{Persuasion and attitude change regarding specific policies are complex and vary depending on the topic. } %When it comes to persuasion and attitude change, it's important to keep in mind that persuasion on specific policies can be a complex issue \revAliFinal{and dependent on the specific topic and policy issue}.
For example, research in social psychology suggests that \revAliFinal{inter-individual} differences can impact attitude change by affecting one or more underlying processes by which variables induce persuasion \cite{brinol2005individual}. In addition, research demonstrates that attitudes are more amenable to change when new information aligns with their original basis, whether cognitive- or affect-based \cite{haddock2019inter}.
In our study, we included a question that focused on attitudes toward the specific drug legalization policy. 
While we observed small shifts in attitudes against drug legalization, \revMiladFinal{it is} worth noting that there are ongoing arguments for and against this policy in policy circles (as discussed in this New Yorker article: \cite{DrugLegalization}). 
\revMiladFinal{Due to the absence of specific claims or evidence for or against drug legalization in our articles, participants had to rely on their interpretations. This underscores the necessity for visualization and article designers to provide clear, direct evidence linking data to policy when advocating for a policy through visualization. This situation presents intriguing research possibilities on the complexities and effectiveness of data visualization in promoting contentious policy agendas.}
%Since our articles did not provide any specific claims or evidence for or against drug legalization, participants had to rely on their own interpretations. If the goal of a visualization is to increase support for a policy, visualization, and article designers will need to provide clear and direct evidence that links the data presented to the proposed policy. This raises interesting questions for future research on the nuances and challenges of using data visualization to advocate for specific policy agendas that are highly debated.
%\revAliFinal{Ideally, future research should study a diverse range of policy issues to shed light on how different policy issues with varying degrees of polarization and personal knowledge might interact with the persuasive power of interactive storytelling.}

\subsection{Differences between belief elicitation techniques}
\label{sec:discussion_elicitation}
We observed several effects of using belief elicitation techniques in the articles. 
First, in both studies, visual elicitation (Draw condition) resulted in \revMiladFinal{a} more accurate recall of the presented data compared to the categorization elicitation technique \revDoug{(Categorize condition)}. 
\revDoug{After drawing a trend, participants may pay more attention to differences between their drawing and the presented data \cite{koonchanok2021data}, consistent with the idea that when combined with data visualization, visual elicitation provides direct feedback to participants \cite{Yea-Seul2017}. This enhanced attention to finer-grained details of the data (e.g., inflection points or short-term fluctuations) may then lead to better recall when asked to reproduce the chart.}
%Kim et al. \cite{Yea-Seul2017} described the visual elicitation as providing direct feedback to participants, and Koonchanok et al. highlighted that people are drawn to notice the discrepancies between their input and the data \cite{koonchanok2021data}. }
In contrast, participants in the Categorize condition were asked to identify the overall trend using loosely defined categories like ``slightly increased" or ``significantly increased."
\revDoug{While the displayed chart still provides feedback, participants' evaluation of the chart may be at the same level of detail, e.g. checking whether the overall trend confirms their selection. This may lead to less scrutiny of specific details and poorer recall later on.} \revDougFinal{One potential confound on drawing a trend leading to better recall compared to categorize a trend is the matching of elicitation and recall technique. It might be interesting to explore in a future study to develop a recall technique that doesn't match any of the elicitation techniques (draw and categorize) yet still captures participants' memory in high fidelity.}
%\revDoug{Categorization} may lead participants to refer to the charts only to confirm their existing belief (e.g., participants are indeed right that the trends did increase) without scrutinizing them closely. 

It is important to note that neither the Draw nor Categorize elicitation conditions led to noticeable differences \revDoug{in recall error} from the Control condition with no elicitation.
\revDoug{One possibility is that the Control condition represents variability in unprompted attention to fine-grained visual details vs. evaluation of the overall direction of a trend, while the Visual and Categorize direct attention in one way or the other. } \revAli{Moreover, we did not control for the impact of y-axis limits on participants' recall errors for each individual chart. Instead, we opted to keep the y-axis limit constant for each question for \revMiladFinal{consistency}. For example, the chart on deaths from HIV (Fig 3 middle) shows a 50 percent decrease but looks flatter than other charts. This might produce some issues with chart recall that should be addressed in future studies.}
%This suggests that, at least in the short term, using elicitation techniques might not provide benefits in terms of recalling specific patterns. 
In the future, it is important to assess whether elicitations have long-lasting effects on recall. 

We also observed that in Study 2, participants, on average, had larger recall errors. 
\revDoug{Although we hesitate to draw strong conclusions about the cause of this difference (see next section),} we suspect it might be caused by cognitive load due to the increased number of elicitations in the articles. We suspect that recollection of trends in real-world settings where consumers rapidly consume information might lead to even worse recollections. It is important to note, based on the findings from our studies, that attitude change is almost independent of recall accuracy. In other words, participants potentially found the articles persuasive enough to update their attitude without being able to recreate the trends they saw in the articles accurately. Our findings raise interesting questions about the affordances and differences between different belief elicitation techniques.

Moving beyond recall, we observe other potential benefits of using visual elicitation techniques. Notably, we observe that in both studies, participants expressed higher levels of surprise in the Draw condition (See figure \ref{fig:study1-engagement}). This is interesting because surprise is known to be an important factor in persuasion and trust \cite{petty2001individual}. One way to explain this response is through the elaboration likelihood model. It is possible that when participants are asked to draw a pattern before being exposed to it, they may elaborate more on the discrepancies and thus report feelings of higher surprise. 

\revAli{Although we do observe impacts of elicitation techniques on surprise, compared to control, we do not observe an effect of elicitation on attitude change. It is possible that the provided narratives and visualizations that are shared between all conditions provide a stronger impact on attitude change for this topic, and the potentially smaller impact of belief elicitation might go unnoticed. In future studies, the impact of elicitation with and without provided narratives might shed light on this point.}

\subsection{The surprisingly effective \textcolor{revision2}{contrasting narratives}}

In Study 1, each article presented only one visualization to directly convey the main message. However, the ``You Draw It" article by the NY Times \cite{YouDrawIt} took a different approach. 
It included two visualizations with relatively flat trends that sharply contrasted the increasing trend of the last visualization, which is related to the article's main topic of the drug epidemic. 
This contrast between the first two datasets and the main visualization might amplify the level of surprise \revDoug{at the rapid recent acceleration of the drug overdose crisis} and lead to \revDoug{stronger feelings of} engagement. 

\revDoug{Our combined analysis of Studies 1 and 2 provide initial evidence that contrasting narratives (Study 2)} have the potential to increase engagement in terms of feelings of surprise and interest \revDoug{compared to when the same data is presented in isolation (Study 1)}.
\revDoug{As we noted earlier, however, our conclusions from this integrative analysis \cite{curran_integrative_2009} are limited by a number of potential confounds. In order to present contrasting narratives in Study 2, participants were necessarily shown a greater number of charts. Heightened engagement could result from presenting additional information separately from the specific nature of the contrast in each series of charts.} \revDougFinal{ In addition, since the studies were run at different timepoints, there might be other differences between the samples which could contribute to the observed effects. In future work, manipulating the content of the charts to control the degree of contrast could better isolate the role of contrasting narratives in driving enhanced surprise and interest. }

\begin{figure}[ht]% specify a combination of t, b, p, or h for top, bottom, on its own page, or here
  \centering % avoid the use of \begin{center}...\end{center} and use \centering instead (more compact)
  \includegraphics[width=\columnwidth]{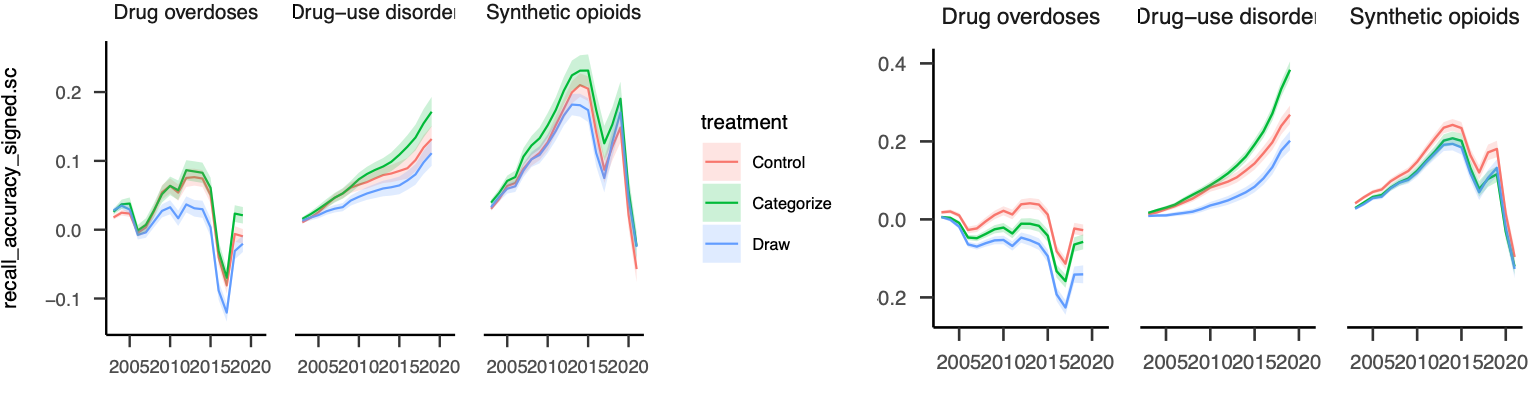}
  \caption{Signed recall error (mean and standard error) in Study 1 \& 2.
  }
  \label{fig:signed_recall_error}
\end{figure}

\revDoug{With those caveats in mind, it is interesting to note that the combined analysis also suggested that participants had poorer memory for the visualizations}
% Our study uncovered several interesting patterns that may be considered unintended consequences on recall 
when adding \textcolor{revision2}{contrasting narratives}. For instance, comparing the average recall results from Study 2 (Figure \ref{fig:recall} right) to Study 1 (Figure \ref{fig:recall} left), we found that participants consistently underestimated the trends on Drug Overdoses and Synthetic Opioids, as shown by the enlarged gaps between the average guessed trend to the ground truth. In contrast, participants systematically overestimated the trend on Drug-user disorders. In other words, participants underestimated the more drastic trends while overestimating the ``milder" trends. Figure \ref{fig:signed_recall_error} shows such systematic under and over estimation with the signed recall error, which is calculated with by subtracting the mean of responses ($\bar{\hat{y}}$) from the mean of true trend ($\bar{y}$ ) for each chart. %had higher recollections of the presented datasets for the drug-use disorder article than the actual ground truth. In other words, it appeared that participants were systematically overestimating the observed pattern. 
 Note that the range of the absolute error (y axis) is larger in Study 2 compared to Study 1. 
This effect may have been caused by the contrast introduced by the first two visualizations in the each article. These unintended consequences warrant further exploration to better understand their underlying causes.
As designers of interactive articles, the choice to orchestrate multiple datasets to convey meaning or elicit a reaction is an interesting and under-explored aspect of visualization research that deserves further exploration.

\section{Conclusion}

Our work is inspired by the ``You Draw It" article published by New York Times \cite{YouDrawIt}. Two techniques we found particularly interesting are eliciting users' prior beliefs expressed by drawing a timeline, and the effective use \textcolor{revision2}{contrasting narratives} to accent the main data visualization on a specific issue. We designed two experiments to evaluate the impact of elicitation and \textcolor{revision2}{contrasting narratives} on attitude change, recall, and engagement. In the first study that focus on belief elicitation, we compared two different forms of belief elicitation with a control condition with no elicitation. Our findings revealed an overall attitude change and that the draw trend elicitation leads to higher engagement in terms of feelings of surprise. In the second study, we evaluate the impact of \textcolor{revision2}{contrasting narratives}. Compared to the results on study 1, we found that \textcolor{revision2}{contrasting narratives} increased attitude change, and improved engagement in terms of surprise and interest, but interestingly resulting in higher recall error. Our discussion further highlights the significance of our findings and many future research opportunities.

\section*{Acknowledgments}
This research is supported by NSF CNS-1747785 and NSF CNS-2323795. The authors also thank anonymous reviewers for their valuable comments.

% {\appendix[Proof of the Zonklar Equations]
% Use $\backslash${\tt{appendix}} if you have a single appendix:
% Do not use $\backslash${\tt{section}} anymore after $\backslash${\tt{appendix}}, only $\backslash${\tt{section*}}.
% If you have multiple appendixes use $\backslash${\tt{appendices}} then use $\backslash${\tt{section}} to start each appendix.
% You must declare a $\backslash${\tt{section}} before using any $\backslash${\tt{subsection}} or using $\backslash${\tt{label}} ($\backslash${\tt{appendices}} by itself
%  starts a section numbered zero.)}

%{\appendices
%\section*{Proof of the First Zonklar Equation}
%Appendix one text goes here.
% You can choose not to have a title for an appendix if you want by leaving the argument blank
%\section*{Proof of the Second Zonklar Equation}
%Appendix two text goes here.}

 % argument is your BibTeX string definitions and bibliography database(s)
%\bibliography{IEEEabrv,../bib/paper}
%

\bibliographystyle{IEEEtran}
\bibliography{template}

% \begin{thebibliography}{1}
% \end{thebibliography}

\newpage

\section{Biography Section}
% If you have an EPS/PDF photo (graphicx package needed), extra braces are
%  needed around the contents of the optional argument to biography to prevent
%  the LaTeX parser from getting confused when it sees the complicated
%  $\backslash${\tt{includegraphics}} command within an optional argument. (You can create
%  your own custom macro containing the $\backslash${\tt{includegraphics}} command to make things
%  simpler here.)
 
%\vspace{11pt}

\begin{IEEEbiography}[{\includegraphics[width=1in,height=1.25in,clip,keepaspectratio]{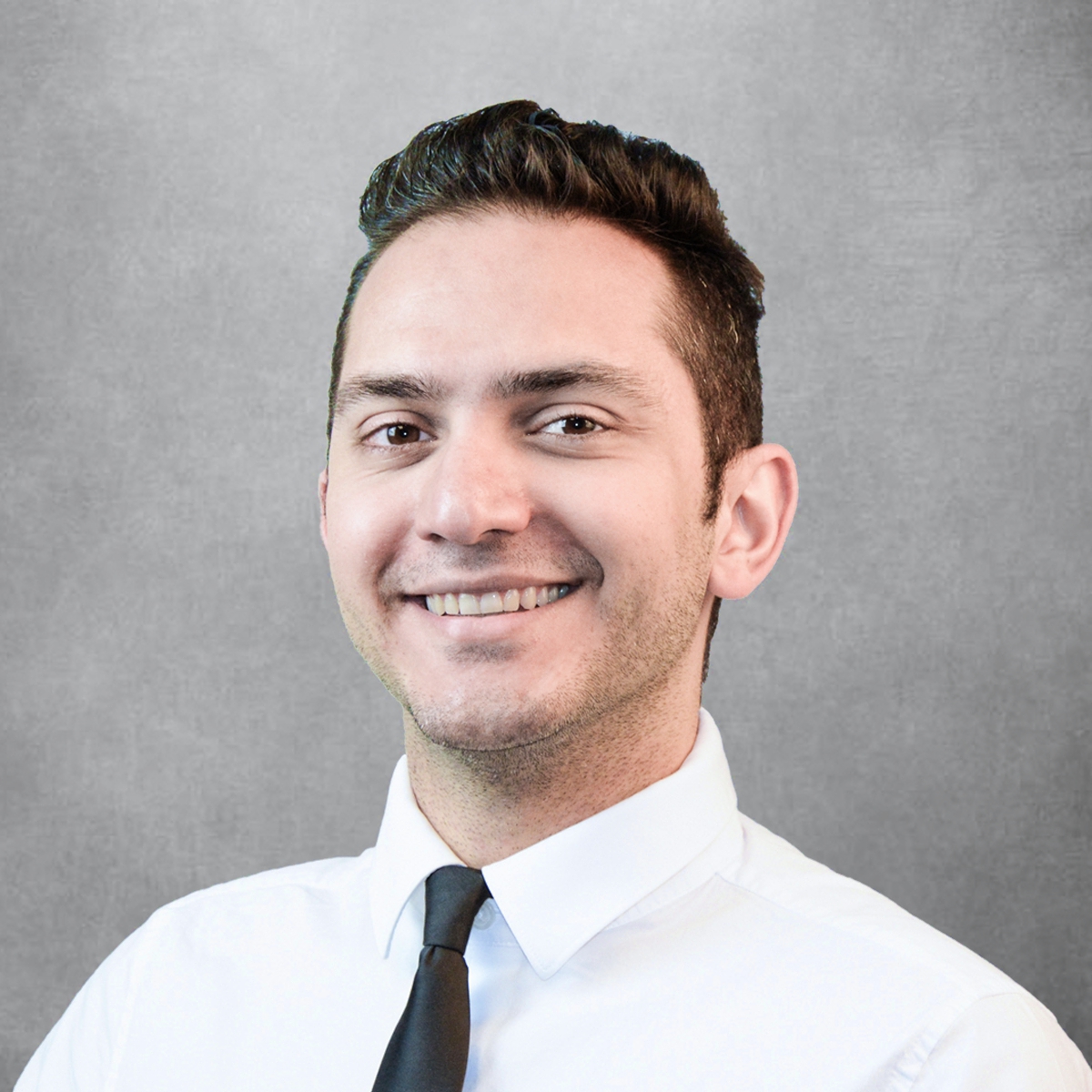}}]{Milad Rogha}
holds a bachelor's degree in Electrical Engineering and a master's degree in Architecture. He is working toward a Ph.D. in Computer Science at the University of North Carolina at Charlotte. His research interests include human-AI interaction and the impact of visualization on human decision-making. His research is dedicated to enhancing artificial intelligence applications that bolster human decision-making processes.
\end{IEEEbiography}
\vspace{5pt}
\begin{IEEEbiography}[{\includegraphics[width=1in,height=1.25in,clip,keepaspectratio]{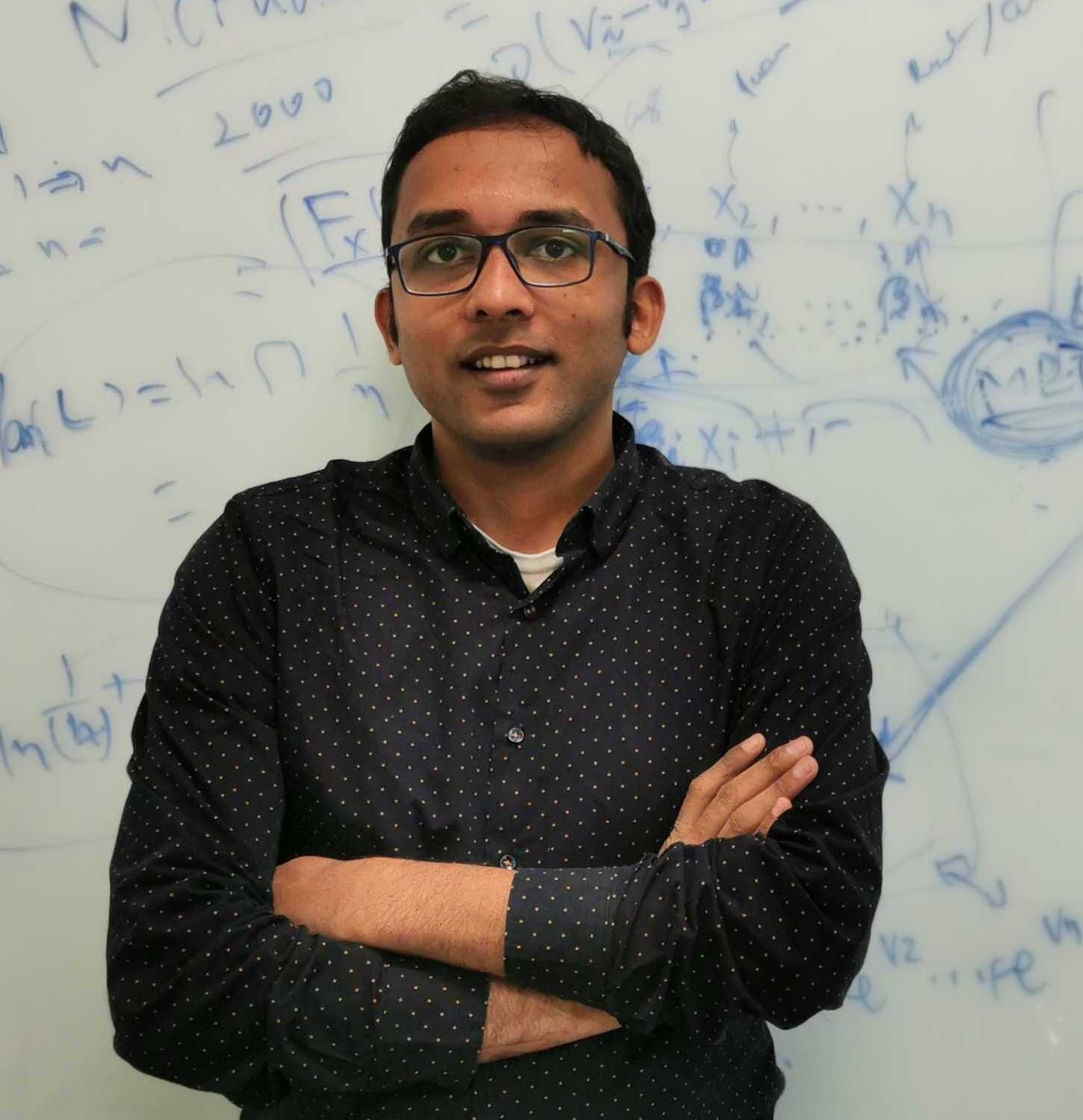}}]{Subham Sah}
 holds a Dual Master's of Science degrees in Software and Information Systems ,and Architecture from the University of North Carolina Charlotte. He loves building products that can engage and educate people. His area of interest is creating Visual Analytics user interfaces that assist users in becoming more conscious of their analytical behaviours and directing them towards  achieving their goals.
\end{IEEEbiography}

\begin{IEEEbiography}[{\includegraphics[width=1in,height=1.25in,clip,keepaspectratio]{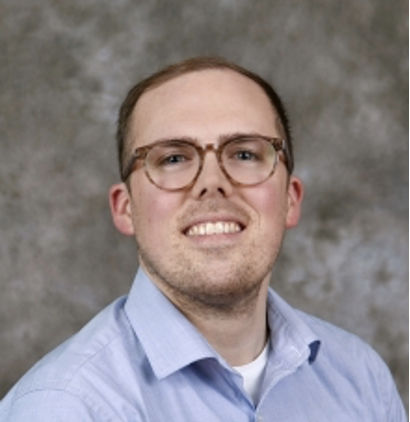}}]{Doug Markant}
is a cognitive scientist who examines human learning, memory, and decision making through behavioral experiments and computational modeling. He is currently an Assistant Professor in the Department of Psychological Science at UNC Charlotte, where he is a core faculty member of the Health Psychology Ph.D. program and an affiliate of the Ribarsky Center for Visual Analytics. He also serves as the director of the interdisciplinary program in Cognitive Science.
\end{IEEEbiography}

\begin{IEEEbiography}[{\includegraphics[width=1in,height=1.25in,clip,keepaspectratio]{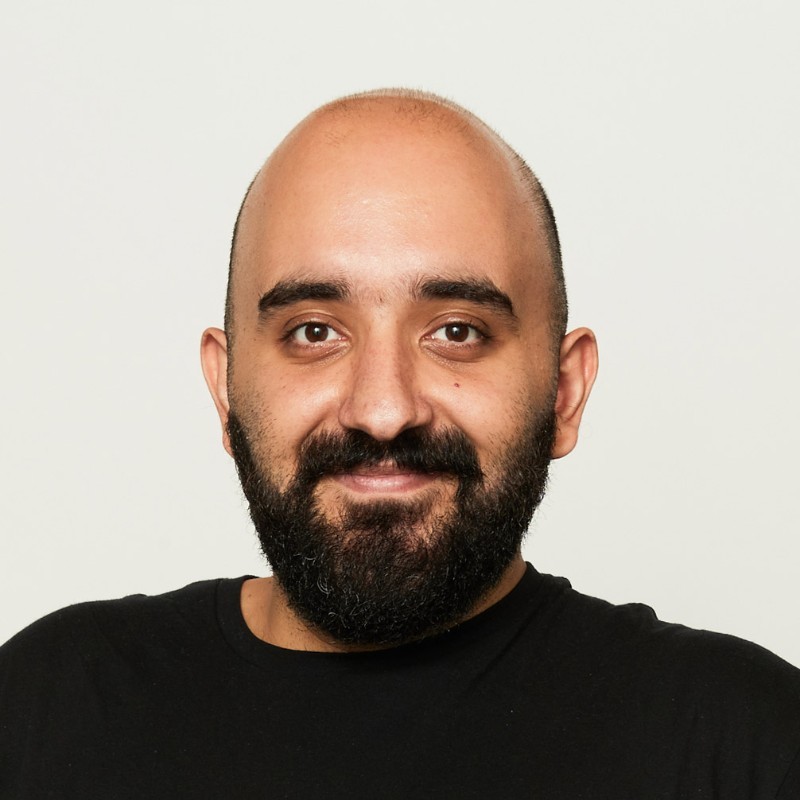}}]{Alireza Karduni}
 is an Assistant Professor of Human-Centred Computing at Simon Fraser University's School of Interactive Arts and Technology. His work focuses on designing interactive visual tools that help people understand and make decisions based on new information. Recognizing the socially and politically situated nature of information ecosystems, he researches how our existing views might influence our receptivity to new (mis)information and how such dynamics might influence societal interactions. 
\end{IEEEbiography}

\begin{IEEEbiography}[{\includegraphics[width=1in,height=1.25in,clip,keepaspectratio]{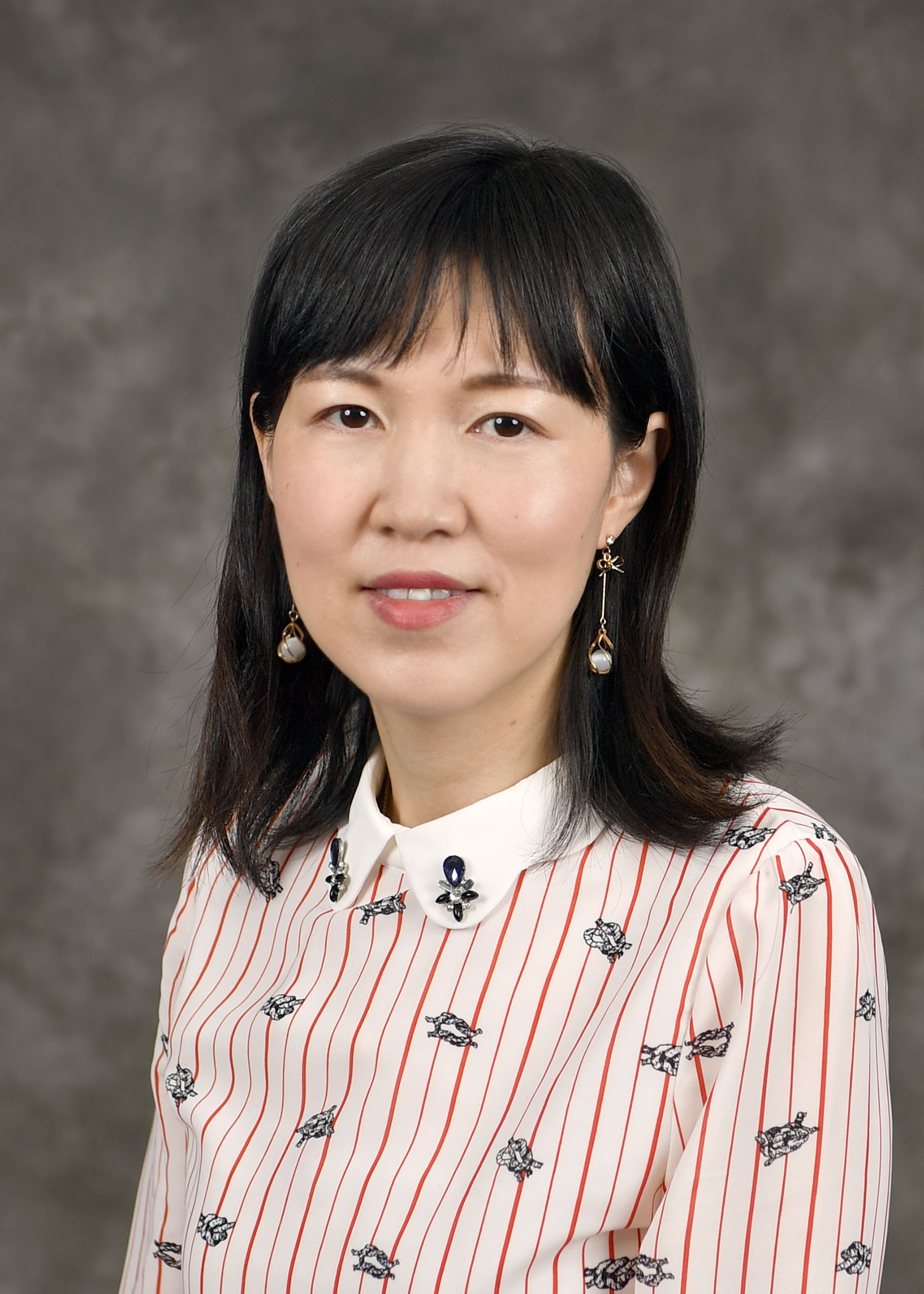}}]{Wenwen Dou}
is an associate professor at the University of North Carolina at Charlotte and co-director of the Ribarsky Center for Visual Analytics. Her research interests include Visual Analytics, Text Mining, and Explainable AI. Dou has worked with various analytics domains in reducing information overload and providing interactive visual means to analyzing unstructured information. She has experience in turning cutting-edge research into technologies that have broad societal impacts.
\end{IEEEbiography}

\vfill

\end{document}